\documentclass{JHEP}
\usepackage{epsf}
\usepackage{psfig}
\usepackage{epsfig}
\usepackage{citesort}

\def\ie{{\it i.e.}}

\def\21{$SU(2) \ot U(1)$}
\def\ot{\otimes}
\def\ie{{\it i.e.}}

\def\bold#1{\setbox0=\hbox{$#1$}
     \kern-.025em\copy0\kern-\wd0
     \kern.05em\copy0\kern-\wd0
     \kern-.025em\raise.0433em\box0 }

\newcommand{\lsim}{\mbox{\raisebox{-1.ex}
{$\stackrel{\textstyle <}{\textstyle \sim}$}}}
\newcommand{\gsim}{\mbox{\raisebox{-1.ex}
{$\stackrel{\textstyle >}{\textstyle \sim}$}}}
\newcommand{\sstt}      {\sin^2 2\theta}
\newcommand{\sinsq}      {\sstt}
\newcommand{\dms}       {\Delta m^2}
\newcommand{\dmsq}       {\dms}
\newcommand{\nue}       {{\nu}_{\rm e}}
\newcommand{\nuebar}    {\bar{\nu}_{\rm e}}
\newcommand{\numu}      {\nu_{\rm \mu}}

\newcommand{\nutau}     {{\nu}_{\rm \tau}}

\newcommand{\units}[1] {\mbox{$\textstyle #1$}}

\def\npb#1#2#3{    { Nucl. Phys. }{\bf B #1} (19#2) #3}
\def\npbps#1#2#3{  { Nucl. Phys. }(Proc. Suppl.){\bf B #1} (19#2) #3}
\def\plb#1#2#3{    { Phys. Lett. }{\bf B #1} (19#2) #3}
\def\prd#1#2#3{    { Phys. Rev. }{\bf D #1} (19#2) #3}

\def\prl#1#2#3{    { Phys. Rev. Lett. }{\bf #1} (19#2) #3}

\def\mpla#1#2#3{   { Mod. Phys. Lett. }{\bf A #1} (19#2) #3}

\def\ibid#1#2#3{   {\it  ibid. }{\bf #1} (19#2) #3}
\title{Neutrino mass and oscillation as probes of physics 
 beyond the Standard Model~\footnote{\it
Invited article prepared for the Journal of the Egyptian Mathematical
Society.}} 
\author{S. Khalil$^{1,2,\dagger}$ and E. Torrente-Lujan$^{3,\dagger}$\\
$^1$ Centre for Theoretical Physics, University of Sussex, Brighton BN1
9QJ,U.K.\\
$^2$ Ain Shams University, Faculty of Science, Cairo
11566, Egypt.\\
$^3$ Departamento de F\'{\i}sica
Te\'orica, C-XI, Universidad Aut\'onoma de Madrid, 
\phantom{$^{4}$} 28049~Cantoblanco, Madrid, Spain.\\
$^\dagger$ E-mail: 
kafz8@pact.cpes.susx.ac.uk, e.torrente@cern.ch.}

\abstract{We present a review of the present status of the problem of
neutrino masses and mixing including  a survey of theoretical
motivations and models, experimental
 searches and implications of recently appeared  solar and
atmospheric neutrino data, which strongly indicate nonzero
neutrino masses a mixing angles. }

\preprint{FTUAM/00-02\\
SUSX-TH-00-023}

\begin{document}

\section{Introduction}

The existence of 
a so-called neutrino, a light, neutral, feebly interacting fermion,  was 
first  proposed by W. Pauli in 1930 
to save the principle
of energy conservation in nuclear  beta decay~\cite{pauli}. 
The idea was promptly adopted by the physics community; 
in 1933 E. Fermi takes the neutrino  hypothesis, gives the neutrino its name  and builds his 
theory of beta decay and weak interactions.
But, it was only 
in 1956 that C. Cowan and F. Reines were able to discover 
the neutrino, more exactly the anti-neutrino,
experimentally~\cite{cowan}. Danby et al.~\cite{danby}
confirmed in 1962 that there exist, at least, two  types of
neutrinos, the $\nu_e$ and $\nu_\mu$. 
In 1989, the study of the Z boson lifetime allows to show with
 great certitude that only three light neutrino species do exist.
Only in 2000, it has been confirmed   by 
direct means \cite{tauneutrino} the existence of the third type of neutrino, 
the $\nu_\tau$ in addition to the $\nu_e$ and $\nu_\mu$.
Until here the history, with the present perspective 
we  can say that
the neutrino occupies a unique place among all
the fundamental particles in many ways and as such it has shed light on
many important aspects of our present understanding of nature and is still
believed to hold a key role  to the physics beyond the Standard Model (SM).

In what respects its mass, 
Pauli initially expected the mass of the neutrino to be small
but not necessary zero:  not very much more than the electron mass, 
 F.  Perrin in 1934 showed that its  mass 
has to be less than that of the electron. 
After more than a half a century, 
the question of whether the neutrino has mass is still one open 
question, being one of the
outstanding issues in  particle physics, astrophysics, cosmology
and theoretical physics in general. 
Presently,
there are several theoretical,
observational and experimental motivations which justify the
searching for possible non-zero neutrino masses (see i.e.
\cite{langacker1,fuk2,vallereview,RAMOND2,wilczek1,bilenky98,bilenky9812}
for excellent older reviews on this matter).

Understanding of 
fermion masses in general are one of the major problems of the
SM and observation of the existence or confirmation of non-existence
 of neutrino masses
could introduce  useful new perspectives on the subject. If they are 
confirmed as 
massless they would be the only fermions with this property. A
property which is not dictated by any known fundamental underlying
principle, such as gauge
invariance in the  case of the photon. If it is concluded  that
 they are massive then the question is
why are their masses so much smaller than those of their charged
partners. Although theory alone can not predict neutrino masses,
it is certainly true that they are strongly suggested by present
theoretical models of elementary particles and most extensions of
the SM definitively require neutrinos to be massive.
They therefore constitute a powerful probe of new physics at a
scale larger than the electroweak scale. 

If massive, the superposition postulates of quantum theory predict that 
neutrinos, particles with identical quantum numbers,
 could oscillate in flavor  space. 
If the absolute difference of masses among them is small enough then these 
oscillations could have important phenomenological consequences.
Some hints at accelerator experiments 
as well as  the observed indications of spectral distortion 
and deficit of solar neutrinos
and the anomalies on the 
ratio of atmospheric $\nu_e/\nu_\mu$ neutrinos and their
zenith distribution are naturally accounted by the oscillations of
a massive neutrino. Recent claims of the high-statistics high-precision 
Super-Kamiokande (SK) experiment are unambiguous and left
little room for the scepticism as we are going to see along this review.

Moreover, neutrinos are basic ingredients of astrophysics and cosmology.
There may be a hot dark matter component (HDM) to the Universe:
simulations of structure formation fit the observations only when
some significant quantity  of HDM is included.
If so, neutrinos would be, at least by weight, 
one of the most important ingredients in the Universe.

Regardless of  mass and oscillations,
astrophysical interest in the neutrino and their properties
arises from the fact that it is copiously
produced  in high temperature and/or high density environment and
it often dominates the physics of those astrophysical objects. The
interactions of the neutrino with matter is so weak that it passes
 freely through any ordinary matter existing in the Universe. This makes
neutrinos to be a very efficient carrier of energy drain from optically
thick objects becoming  very  good probes for the interior of
such  objects.
For example, the solar neutrino flux is, together with heliosysmology,
 one of the two known probes of the solar core. A similar statement
 applies to objects as  the type-II supernovas: the most interesting questions
around supernovas, the explosion dynamics itself with the
shock revival, and,  the synthesis of the heaviest elements by
the so-called  r-processes, could be positively affected by changes in the neutrino flux,
e.g. by MSW active or sterile conversions \cite{supernova}. Finally, ultra
high energy neutrinos are called to be useful probes of diverse distant
astrophysical objects. Active Galactic Nuclei (AGN) should be copious
 emitters of $\nu$'s, providing both detectable point sources and an
 observable diffuse background which is larger in fact than the atmospheric
 neutrino background in the very high energy range \cite{AGN}.

This review is organized as follows, in section 2 we discuss the neutrino
in the SM. Section 3 is devoted to the possible ways for
generating neutrino mass terms and different models for these possibilities
are presented. Neutrino oscillation in vacuum and in matter are
studied in section 4. The cosmological and the astrophysical constraints
on diverse
neutrino properties are summarized
 in section 5.
In section 6 we give an introduction to the phenomenological 
description of neutrino oscillations in vacuum and in matter. 
In section 7 we give an extensive
description of the different neutrino experiments, their 
results and their interpretation.
Finally we present  some  conclusions and final remarks 
in section 7.

\section{The neutrino in the Standard Model.}

The current Standard Model of particles and interactions 
 supposes the existence of three  neutrinos. 
The three neutrinos are
represented by two-component Weyl spinors
each describing a left-handed  fermion.
They are the neutral, upper components of 
doublets $L_i$ with respect the  $SU(2)$ group, the 
weak interaction group, we have,
$$
L_i\equiv \left(\begin{array}{c} \nu_{i} \\ l_i
\end{array}
\right), \hspace{1.2cm} i = (e, \mu, \tau).
$$
They 
have the third component of the weak isospin 
$I_{3W}=1/2$ and are assigned an unit of the global $i$th lepton number.
The three right-handed charged  leptons have however no 
counterparts in the neutrino sector and transform as singlets
 with respect the weak interaction.

These SM neutrinos are strictly massless, the reason for this 
 can be understood as follows.
The only
Lorenz scalar made out of them is the Majorana mass, of the form
$\nu^t_i \nu_i$; it has the quantum number of a weak isotriplet, with
$I_{3W}=1$ as well as two units of total lepton number. Thus to generate
a renormalizable
Majorana mass term at the tree level one needs a Higgs isotriplet
 with two units of lepton number. Since in the stricter version of the 
 SM the 
 Higgs sector is only constituted by a weak isodoublet, there are 
 no tree-level neutrino masses.
When  quantum corrections  are introduced we should consider
 effective terms where a weak isotriplet is made out of two isodoublets and
 which are not invariant under lepton number symmetry. 
The conclusion is that in the
SM neutrinos are kept massless by a global chiral lepton number
 symmetry
(and more general properties as renormalizability of the theory, 
see Ref.\cite{RAMOND2} for an applied version of this argument).
However this is a rather formal conclusion, there is no any other 
independent, compelling theoretical
argument in favor of such symmetry, or, with other words,
 there is no reason why we would like to
keep it intact. 

Independent from mass and charge oddities, in any other respect
neutrinos are very well behaved particles within the SM framework and 
some figures and facts are unambiguously known about them.
The LEP  Z boson line-shape measurements imply that are
only three ordinary (weak interacting) light neutrinos \cite{SM,PDG98}.
Big Bang Nucleosynthesis (BBN) constrains the parameters of
possible sterile  neutrinos, non-weak interacting or those
which interact and are produced only by mixing \cite{BBN}.
{\em All the existing} data on  the weak interaction
processes in which neutrinos take part are perfectly described by the
SM
charged-current (CC) and neutral-current (NC) Lagrangians:
\begin{eqnarray}
L_I^{CC}&=&-\frac{g}{\surd 2} \sum_{i=e,\mu,\tau}
\overline{\nu_{L}}_i\gamma_\alpha {l_{L}}_i W^\alpha+ h.c.\\
L_I^{NC}&=&-\frac{g}{2 \cos \theta_W} \sum_{i=e,\mu,\tau}
\overline{\nu_{L}}_i\gamma_\alpha {\nu_{L}}_i Z^\alpha + h.c.
\end{eqnarray}
where $Z^\alpha,W^\alpha$ are the neutral and charged vector bosons 
intermediaries of the weak interaction. 
The CC and NC interaction Lagrangians  conserve 
 three total 
additive quantum numbers, the
lepton numbers $L_{e,\mu,\tau}$ while the structure of the 
CC interactions  is what determine 
the notion of flavor neutrinos $\nu_{e,\mu,\tau}$.

There are no indications in favor of the 
violation of the conservation of these lepton numbers 
in weak
 processes and  very strong bounds
on branching ratios of rare, lepton number violating,
 processes are obtained, for
examples see  Table~\ref{tttt1}.

\begin{table}[h]
$$\begin{array}{|lcc|lcc|}\hline
R(\mu\to e\gamma) &<& 4.9\times 10^{-11} 
  & R(\tau\to e\gamma)&<& 2.7\times 10^{-6}\\
R(\mu\to 3 e) &<& 1.0\times 10^{-12} 
  & R(\tau\to \mu\gamma)&<& 3.0\times 10^{-6}\\
R(\mu\to  e(2\gamma))&<& 7.2\times 10^{-11} 
  & R(\mu\to 3 e)&<& 2.9\times 10^{-6}.\\ \hline 
\end{array}
$$
\caption{Some lepton number violating processes. See 
Ref.\protect\cite{PDG98}, all limits at 90\% CL.}
\label{tttt1}
\end{table}

{}From the theoretical point of view, 
in the minimal extension of the SM where 
right-handed neutrinos are introduced and the neutrino gets a mass,
 the branching ratio of the 
$\mu\to e \gamma$ 
decay is given by (2 generations are 
 assumed \cite{muegamma}),
\begin{eqnarray}
R(\mu\to e\gamma) & =& G_F \left (\frac{\sin 2\theta\  \Delta m_{1,2}^2}{2 M_W^2}\right )^2
\nonumber
\end{eqnarray}
where  $m_{1,2}$ are the neutrino masses, $M_W$ is the mass 
of the $W$ boson and $\theta$ is the mixing angle in the lepton
 sector. Using the experimental upper limit on the 
heaviest $\nu_\tau$ neutrino one obtains $R\sim 10^{-18}$, 
 a value far from being measurable at present as we can see from 
table~\ref{tttt1}
The 
 $\mu\to e \gamma$ 
and similar processes are sensitive to new particles not 
contained in the SM. The value is highly model dependent 
 and could change by several orders of magnitude if we 
 modify the neutrino sector for example introducing 
an extra number of heavy  neutrinos.

\section{Neutrino mass terms and models.}

\subsection{Model independent neutrino mass terms}

Phenomenologically, Lagrangian mass terms can be viewed as  terms 
describing transitions between 
right (R) and left (L)-handed states.
For a given minimal, Lorenz invariant, 
set of four fields: $\psi_L,\psi_R,(\psi^c)_L,(\psi^c)_R$, would-be
 components of a generic Dirac Spinor, the
most general mass part of the  Lagrangian can be
written as:
\begin{eqnarray}
L_{mass}&=&
m_D \left ( \overline{\psi}_L \psi_R\right )
+\frac{1}{2} m_T \left ( \overline{(\psi_L)^c} \psi_L\right )
+\frac{1}{2} m_S \left ( \overline{(\psi_R)^c} \psi_R\right )+h.c.
\label{e2001}
\end{eqnarray}
In terms of the  newly defined  Majorana fields ($\nu^c=\nu,N^c=N$): 
$\nu=(1/\sqrt 2)
(\psi_L+(\psi_L)^c)$, $N=(1/\sqrt 2) (\psi_R+(\psi_R)^c)$, the
Lagrangian  $L_{mass}$ can be rewritten as:
\begin{eqnarray}
L_{mass}&=& \pmatrix{ \overline{\nu} ,& \overline{N} } M
\pmatrix{ \nu \cr N}
\label{e2003}
\end{eqnarray}
where $M$ is the neutrino mass matrix defined as:
\begin{eqnarray}
M&\equiv& 
\pmatrix{ m_T & m_D \cr  m_D & m_S }.
\label{e2003b}
\end{eqnarray}
We proceed further and  
diagonalizing the matrix M one finds that the physical
particle content is given by two Majorana mass
eigenstates: the inclusion of the Majorana mass splits the four
degenerate states of the Dirac field into two non-degenerate Majorana
 pairs.

 If we assume that the states $\nu,N$
are respectively active (belonging to weak doublets) and sterile (weak singlets), the terms corresponding to the ''Majorana masses'' $m_T$ and
$m_S$ transform as weak triplets and singlets respectively. While the
term corresponding to $m_D$ is an standard, weak singlet in most cases, Dirac mass term.

The neutrino  mass matrix can easily be generalized to three
or more families, in which
case the masses become matrices themselves.
The complete flavor mixing comes from two different
parts, the diagonalization of
 the charged lepton Yukawa couplings and that of the neutrino masses.
In most simple extensions of the SM, this CKM-like leptonic
 mixing  is totally
arbitrary with parameters only to be determined by experiment. 
Their  prediction, as for the quark hierarchies and mixing, 
needs  further theoretical assumptions 
(i.e. Ref.\cite{RAMOND,RAMOND2} predicting $\nu_\mu-\nu_\tau$ maximal mixing).

We can analyze different cases.
In the case of  a purely Dirac mass term,  
$m_T=m_S= 0$ in Eq.(\ref{e2003}),
the $\nu,N$ states are degenerate with mass $m_D$ and a  four component
Dirac field can be  recovered as $\nu\equiv \nu+N$.
It can be seen that, although violating individual lepton numbers, 
the Dirac mass term allows a conserved lepton number $L=L_\nu+L_N$.

In the general case,
  pure Majorana mass transition terms,
$m_T$ or $m_S$ terms in  Lagrangian (\ref{e2003}), describe in fact
 a particle-antiparticle transition
 violating lepton number by two units
($\Delta L=\pm 2$). They can be viewed as the creation or annihilation of
two neutrinos leading therefore to the possibility of the 
 existence of  neutrinoless double beta decay.

In the general case where all classes of terms are allowed, it is interesting 
 to consider the so-called
''see-saw'' limit in Eq.(\ref{e2003}). In this limit taking $m_T\sim 1/m_S \sim 0, m_D<< m_S$,   the
two Majorana neutrinos acquire respectively masses
$m_1  \sim  m_D^2/m_S<< m_D$,$ m_2  \sim  m_S$.
There is one heavy neutrino and one neutrino much lighter than the typical
 Dirac fermion mass. 
One of  neutrino mass has been automatically suppressed, balanced up
(``see-saw'')  by the heavy one. 
The ''see-saw'' mechanism is a natural 
way of generating two well
 separated  mass scales.

\subsection{Neutrino mass models}

Any fully satisfactory
model that generates neutrino masses must contain a natural
mechanism that explains their small value, relative to that of their
charged partners.
Given the latest experimental indications it would also  be desirable  that
includes any comprehensive 
justification for light sterile neutrinos and 
large, near maximal, mixing.

Different models can be distinguished 
 according to the new particle content or according to the scale.
According to the particle content,
of the different open possibilities, if we want 
to break lepton number and 
to generate neutrino masses without introducing new fermions
in the SM, we must do it  by  adding to the
SM Higgs sector  fields  carrying lepton numbers, one can
arrange then to break lepton number
explicitly or spontaneously through their interactions.
But, possibly, the most straightforward  approach
 to generate  neutrino masses is to introduce for
each neutrino  an  additional weak neutral singlet.

This happens naturally in the framework of 
LR symmetric models where the origin of
SM parity ($P$) violation is ascribed to the spontaneous breaking of 
 a baryon-lepton (the $B-L$ quantum number)   symmetry.

In the $SO(10)$ GUT the Majorana neutral particle  $N$
enters in a natural way in order to complete the
matter multiplet,
the neutral $N$ is a $SU(3)\times SU(2)\times U(1) $ singlet.

 According to the scale where the new physics
 have relevant effects, Unification
(i.e. the aforementioned $SO(10)$ GUT) and
weak-scale approaches (i.e. radiative models) are usually  distinguished
\cite{varios,varios2}.

The anomalies observed in the solar neutrino flux, atmospheric flux and
low energy accelerator experiments cannot all be explained
consistently without introducing a light, then necessarily sterile,
 neutrino.
If all the Majorana masses are small, active neutrinos
can oscillate into the sterile right handed fields.
Light sterile neutrinos can appear in particular see-saw mechanisms
if additional assumptions are considered
(``singular see-saw `` models) with some unavoidable fine tuning.
The alternative to such fine tuning would be seesaw-like suppression
for sterile neutrinos  involving  new unknown
 interactions, i.e. family symmetries,
 resulting in substantial additions to the SM,
(i.e. some sophisticated superstring-inspired models, Ref.\cite{benakli}).

Finally some example of weak scale models, radiative generated mass
models where the neutrino masses are zero at tree level
constitute a very different class of models: they explain
in principle the smallness of $m_\nu$ for both active and sterile
neutrinos. Different mass scales are generated naturally by different
number of loops involved in generating each of them.
The actual implementation
generally  requires however the ad-hoc introduction of new Higgs
 particles  with nonstandard electroweak quantum numbers and lepton
 number-violating couplings \cite{valle2}.

The origin of the different Dirac and Majorana mass terms $m_D,m_S,M_T$ appearing 
 above is usually  understood by a dynamical mechanism where at some 
scale or another some  symmetry is  spontaneously broken as follows.

First we will deal with the  Dirac mass term.
For the case of interest,  $\nu_L$  and $\nu_R$ are 
 SU(2) doublets and  singlets respectively,
the mass term describes then a $\Delta I=1/2$ transition and
 is generated from  SU(2) breaking with a Yukawa coupling:
\begin{eqnarray}
L_{Yuk}&=&h_i \left (\overline{\nu}_i ,\overline{l_i}\right )_L
\left(\begin{array}{c}
     \phi^0\\
      \phi^- \end{array}\right)  {N_R}_i + h.c. 
\label{e2005}
\end{eqnarray}
Where $\phi_0,\phi¯$ are the components of the Higgs doublet.
The coefficient $h_i$ is the  Yukawa coupling.
One has that, after symmetry breaking, 
$m_D=h_i v/2$ where $v$ is the vacuum expectation value of the
Higgs doublet.
A neutrino Dirac mass is qualitatively just like any other fermion
masses, but that leads
to the question of why it is so small in comparison with the
rest of fermion masses: one would require
 $h_{\nu_e}<10^{-10}$ in order to have $m_{\nu e}< 10 $ eV.
Or in other words: $h_{\nu_e}/h_e\sim 10^{-5}$ while
 for the hadronic sector we have
 $h_{up}/h_{down}\sim O(1)$.

Now  we will deal with  the Majorana mass terms. The 
$m_S$ term will appear if 
 $N$ is  a gauge singlet. In this case 
a renormalizable mass term of the type $L_N=m_S N^t N$ is
allowed by the SM  gauge group $SU(3)\times SU(2)\times U(1)$ symmetry.
However, it would not be consistent in general with
unified symmetries, i.e. with a full SO(10) symmetry and some complicated
 mechanism should be invocated.
A $m_S$ term is usually  associated
 with the breaking of some larger  symmetry, the expected scale for it should be
 in a  range covering from $\sim$ TeV (LR models) to
GUT scales $\sim 10^{15}-10^{17}$ GeV.

Finally,  the $m_T$ term will appear if 
 $\nu_L$ is active, belongs to some gauge doublet. In this case we have 
$\Delta I$=1 and $m_T$ must be generated by either
a) an elementary Higgs triplet or b) by an effective operator involving two
Higgs doublets arranged to transform as a triplet.
In case a),
for an elementary triplet $m_T\sim h_T v_T$, where $h_T$ is a Yukawa
 coupling and $v_T$ is the triplet VEV.
The simplest implementation (the old  Gelmini-Roncadelli model \cite{gelmini})
 is excluded by the LEP data on the Z width: the corresponding
Majoron couples to the Z boson increasing significantly its width.
Variant models involving
explicit lepton number violation or in which the Majoron is mainly a weak
 singlet (\cite{chika}, invisible Majoron models) could still be possible.
In case b), 
for an effective operator originated mass, one expects
$m_T\sim 1/M$ where  $M$ is the scale
of the new physics which generates the operator.

A few words about the range of expected neutrino masses for 
different types  of models depending on the values of $m_D,M_{S,T}$.
For $m_S\sim 1$ TeV (LR models)
and with  typical $m_D$'s, one expects masses of order
 $10^{-1}$ eV, 10 keV, and 1 MeV for the
$\nu_{e,\mu,\tau}$ respectively.
GUT theories motivates a big range of 
intermediate scales $10^{12}-10^{16}$ GeV.
In the lower end of this range,
for $m_S\sim 10^{12}$ GeV
(some superstring-inspired models, GUT with
multiple breaking stages) one can obtain light neutrino masses of the order
$(10^{-7}$ eV, $10^{-3}$ eV, 10 eV). 
At the upper end,
for $m_S\sim 10^{16}$ (grand unified seesaw with large Higgs representations)
one typically finds smaller masses around
$(10^{-11}$, $10^{-7}$, $10^{-2}$) eV
somehow more difficult to fit into the present known experimental facts.

\subsection{The magnetic dipole moment and neutrino masses}

The magnetic dipole moment  is another
probe of possible new interactions.
Majorana neutrinos
have identically zero magnetic and electric dipole moments.
Flavor transition magnetic moments are  allowed however in general for both
Dirac and Majorana neutrinos.
Limits obtained
 from laboratory experiments are of the order of a few
$\times 10^{-10}\mu_B$ and those from stellar physics or cosmology are
$O(10^{-11}-10^{-13})\mu_B$.
 In the SM electroweak theory, extended to allow for Dirac neutrino masses,
the neutrino magnetic dipole moment is nonzero and given, as
(\cite{PDG98} and references therein):
\begin{eqnarray}
\mu_\nu&=& \frac{3 e G_F m_\nu}{8 \pi^2 \surd 2}= 3\times 10^{-19} (m_\nu/1\ eV)\mu_B
\end{eqnarray}
where $\mu_B$ is the Bohr magneton.
The proportionality of $\mu_\nu$ to the
neutrino mass is due to the absence of any interaction of $\nu_R$ other
than its Yukawa coupling which generates its mass. In LR
symmetric theories $\mu_\nu$ is proportional to the charged
lepton mass: a value  of $\mu_\nu\sim 10^{-13}-10^{-14}\mu_B$ can be reached
still too small to have practical astrophysical consequences.

Magnetic moment interactions arise in any renormalizable gauge theory
only as finite  radiative corrections. The diagrams which generate a magnetic
moment will also contribute to the neutrino mass once the external
photon line is removed.
In the absence of additional symmetries a large magnetic moment is
incompatible with a small neutrino mass.
The way out suggested by Voloshin  consists in defining a
SU(2)$_\nu$ symmetry acting on the space $(\nu,\nu^c)$, magnetic moment
terms are singlets under this symmetry. In the limit of
exact SU(2)$_\nu$ the neutrino mass is forbidden but $\mu_\nu$ is
allowed \cite{fukugita}. Diverse concrete models have been
proposed where such symmetry  is embedded into an
extension of the SM (left-right symmetries,  SUSY with horizontal gauge
symmetries \cite{babu1}).

\section{Aspects of some theoretical models for neutrino mass}
\subsection{Neutrino masses in  LR models}

 A very natural way to generate neutrino mass is 
to minimally extend the SM including additional 2-spinors as
right handed neutrinos
and at the same time extend the, non-QCD, SM gauge symmetry group to 
$G_{LR}\equiv SU(2)_L \times SU(2)_R\times U(1)_{B-L}\times P $. 
The resulting model, initially  proposed in 1973-1974, 
is known as the left-right (LR) symmetric model~\cite{mohapatra}. 
This kind of models were  first proposed with the goal of 
seeking a spontaneous origin for $P$ violation in weak interactions: 
 CP and P  are conserved at 
large energies; at low energies, however, the group $G_{LR}$ 
breaks down spontaneously at a scale $M_R$. Any new physics correction
 to the SM would be of order $(M_L/M_R)^2$ where $M_L\sim m_W$; 
if we choose $M_R>>M_L$ we obtain only small corrections, compatible
 with present known physics. We can satisfactorily  
explain in this case 
the small quantity of 
CP violation observed in present experiments and why 
the neutrino mass is so small, as we will see below.

The quarks ($Q$) and leptons  ($L$) in  LR models transform as 
doublets under the  group $ SU(2)_{L,R}$ as follows:
$Q_L,L_L\sim (2,1)$ and
$Q_R,L_R\sim (1,2)$. The gauge interactions are symmetric between
  left and right -handed fermions; therefore  before symmetry 
 spontaneous breaking, weak interactions, as the others,
 conserve parity 

The breaking of the gauge symmetry is implemented by 
  multiplets of  LR symmetric Higgs fields,
the concrete choosing of these multiplets is not unique. It has been shown that 
in order to understand the smallness of the neutrino mass, it is convenient
to choose respectively one doublet and two triplets as follows:
\begin{eqnarray*}
\phi &\sim& (2,2,0)\\ 
\Delta_L \sim  (3,1,2)& ,& \Delta_R \sim (1,3,2).
\end{eqnarray*}
The Yukawa couplings of these Higgs fields to the quarks and leptons are given by
\begin{eqnarray}
L_{yuk} &=& h_1 \bar{L}_L \phi L_R + h_2 \bar{L}_L
\tilde{\phi} L_R +  h_1' \bar{Q}_L \phi Q_R + h_2' \bar{Q}_L
\tilde{\phi} Q_R \nonumber\\
    &+& f(L_L L_L \Delta_L + L_R L_R \Delta_R) + h.c.
\end{eqnarray}

The gauge symmetry breaking proceeds in two steps.
The $SU(2)_R\times U(1)_{B-L}$ is broken down to $U(1)_Y$ by choosing
$\langle \Delta_R^0 \rangle= v_R \neq 0$ since this carries both $SU(2)_R$
and $U(1)_{B-L}$ quantum
numbers. It gives mass to charged and neutral right handed gauge
bosons, \ie,
$$M_{W_R} = g v_R,\quad  M_{Z'} = \sqrt{2} g v_R/\sqrt{1-\tan^2 \theta_W}.$$ 
Furthermore, as consequence of 
$f$-term in the Lagrangian above this stage of symmetry breaking
also leads to a mass term  for the right-handed neutrinos of the order $\sim f v_R$. 

Next, as we break the SM symmetry by turning on
the vev's for $\phi$ fields as 
$ \langle \phi \rangle =diag(v_\kappa, v_\kappa')$, 
with $v_R >> v_\kappa'>> v_\kappa$, 
we give masses to the $W_L$ and the $Z$ bosons and  
also to  quarks and leptons ($m_e\sim h v_\kappa$). 
At the end of the process of spontaneous symmetry breaking the 
 two $W$ bosons of the model will mix, the lowest physical mass 
 eigenstate is identified as the observed W boson.
Current experimental limits set the limit 
(see Ref.\cite{PDG98}, at 90\% CL) $m_{WR}> 550 $ GeV.

In the neutrino
sector the above Yukawa couplings after $SU(2)_L$ breaking by
$\langle \phi \rangle \neq 0$ leads to the Dirac masses for the
neutrino.
The full process  leads to the following mass
matrix for the $\nu$, $N$,
(the matrix M in eq.\ref{e2003b})
\begin{equation}
M = \left( \begin{array}{cc}
         \sim 0 & h v_\kappa \\
        h v_\kappa & f v_R
        \end{array}\right).
\end{equation}
{}From the structure of this matrix
 we can see the {\em see-saw} mechanism 
at work.
By diagonalizing M, we get a light neutrino
corresponding to the eigenvalue 
 $m_{\nu}\simeq (hv_\kappa)^2/f v_R$ and a 
heavy one with mass $m_N\simeq f v_R$.

Variants of the basic $LR$ model include the possibility of 
 having Dirac neutrinos as the expense of enlarging the particle
 content. The introduction of two new singlet fermions 
  and a new set of carefully-chosen 
Higgs bosons, allows us to
 write the $4\times 4$ mass matrix \cite{mohapatrareview}:
\begin{eqnarray}
M&=&\pmatrix{     
 0 &m_D &0 &0  \cr
 m_D &0 &0 &f v_R  \cr
 0 &0 &0 &\mu  \cr
 0 &f v_R &\mu &0  }.
\label{matrix00}
\end{eqnarray}
 Matrix \ref{matrix00}  leads to two Dirac neutrinos, one 
 heavy with mass $\sim f v_R$ and another light with
 mass $m_\nu\sim m_D \mu /f v_R$. This light four component 
spinor  has the correct weak interaction properties to be 
 identified as the neutrino.
A variant of this   model  can be constructed by 
addition of singlet quarks and leptons. One can arrange these 
 new particles  in order that the Dirac mass of the neutrino 
 vanishes at the tree level and arises at the one-loop level via
 $W_L-W_R$ mixing.

Left-right symmetric models can be embedded in
 grand unification  groups. 
The simplest GUT model that leads by successive stages of 
 symmetry breaking  to left-right symmetric models
at low energies is the $SO(10)$-based model. 
A example of LR embedding GUT 
 Supersymmetric theories will be discussed below in the 
 context of Superstring-inspired models.

\subsection{SUSY models: Neutrino masses without right-handed neutrinos}

Supersymmetry (SUSY) models with explicit broken
$R$-parity provide an interesting example of how we can  
 generate neutrino masses without using a
right-handed neutrino but incorporating new  particles and 
enlarging the Higgs sector. 

In a generic SUSY model, due to the
Higgs and lepton doublet superfields have the same $SU(3)_c\times
SU(2)_L \times U(1)_Y$ quantum numbers, we have in the 
 superpotential terms, bilinear or trilinear in the superfields, that  violate baryon and lepton number explicitly. They lead 
 to a  mass for the neutrino but also to
to proton decay with  unacceptable high rates. 
One radical possibility is to introduce by hand a symmetry 
that  rule out these terms, 
this is the role of the $R$-symmetry introduced in the MSSM.


A less radical possibility is to allow for the existence in the 
 of superpotential of  
a bilinear term, i.e. $W = \epsilon_3 L_3 H_2$.
This is  
simplest way to illustrate the idea of generating neutrino
mass 
without spoiling current limits on proton decay.
The bilinear violation of
$R$-parity implied by  the  $\epsilon_3$ term 
leads~\cite{valle}  by a minimization condition 
to a non-zero sneutrino vev, $v_3$.
In such a model the $\tau$  neutrino acquire a mass, due to the mixing
between neutrinos and neutralinos. 
The $\nu_{e}$ and $\nu_{\mu}$ neutrinos 
remain massless in this model, it is supposed that 
they get masses from scalar loop contributions. 
The model is phenomenologically equivalent to a three Higgs 
 doublet model where one of these doublets (the 
 sneutrino) carry  a lepton 
number which is broken spontaneously.
We have the following mass matrix for the 
neutralino-neutrino sector, in block form the $5\times 5$ matrix reads:
\begin{equation}
M = \left[ \begin{array}{c|c}
        G & Q \vphantom{Q^t} \\[0.1cm] \hline
        Q^t &
         \begin{array}{ccc}
         0 & -\mu & 0\\
         -\mu & 0 &\epsilon_3\\
         0 & \epsilon_3 & 0
        \end{array} 
	   \end{array}
\right] 
\label{matrix}
\end{equation}
where $G=diag(M_1,M_2)$ corresponding 
to the  two gauginos masses. 
The $ Q$ is a $2\times3$ matrix containing
 $v_{u,d,3}$ 
the vevs of $H_1$, $H_2$ and the sneutrino.
The 
next two rows are Higgsinos
and the last one denotes the tau neutrino. 
Let us remind that gauginos and Higgsinos are the 
supersymmetric fermionic counterparts of the gauge and 
Higgs fields. 

In diagonalizing the mass matrix $M$, 
a ``see-saw'' mechanism is again at work, 
in which the role of $M_D, M_R$ scale masses
are easily recognized. 
It turns out that $\nu_{\tau}$ mass is given by
($v_3'\equiv \epsilon_3 v_d + \mu v_3$), 
$$m_{\nu_\tau}\propto \frac{(v_3')^2}{M},$$ 
where $M$ is the largest gaugino mass.
However, in an arbitrary SUSY model this
mechanism leads to (although relatively small if $M$ is large) still 
too large $\nu_{\tau}$ masses. To obtain a realistically 
small $\nu_{\tau}$ mass we have to assume universality among the
soft SUSY breaking terms at GUT scale. In this case the
$\nu_{\tau}$ mass is predicted to be small due to a cancellation
between the two terms which makes negligible the 
$v_3'$.

%


We consider now the properties of neutrinos in superstring
models. In a number of 
these models, the effective theory 
 imply a supersymmetric $E_6$ grand unified
model, with  matter fields 
belonging to the $27$  dimensional representations
of $E_6$ group plus additional 
$E_6$-singlet fields. The model contains  additional  neutral leptons in 
 each generation and  neutral $E_6$-singlets, 
gauginos and Higgsinos. 
As before but with a larger number of them, 
all of these neutral particles can mix, 
making the understanding of neutrino
 masses quite difficult if no simplifying assumptions are employed.

Several of these mechanisms have been proposed to understand 
neutrino masses \cite{mohapatrareview}.
In some  of these mechanisms 
the huge neutral mixing mass matrix is reduced drastically down to 
a $3\times 3$ neutrino
 mass matrix result of the mixing of the $\nu,\nu^c$ with an 
 additional neutral field $T$ whose nature depends on the 
particular mechanism. 
In the basis $(\nu, \nu^c, T)$
the mass matrix is of the form (with $\mu$ possibly being zero):
\begin{equation}
M = \left( \begin{array}{ccc}
        0 & m_D & 0 \\
        m_D & 0 & \lambda_2 v_R\\
        0 & \lambda_2 v_R & \mu
        \end{array} \right).
\label{matrix2}
\end{equation}
We distinguish two important cases, the R-parity violating 
case and the mixing with a singlet,
where the sneutrinos, superpartners of $\nu^c$, are assumed 
 to acquire a v.e.v. of order $v_R$.

In the first case the $T$ field corresponds to 
  a gaugino with a Majorana mass $\mu$
  that can arise at two-loop order. 
Usually 
$\mu \simeq 100$ GeV, if we 
assume  $\lambda v_R\simeq 1$ TeV 
additional dangerous mixing with the
 Higgsinos can be neglected and 
we are  lead to
 a neutrino mass $m_\nu\simeq 10^{-1}$ eV. Thus, smallness
of neutrino mass is understood without any fine tuning of parameters.

In the second case the field $T$ corresponds to 
one of the $E_6$-singlets presents in the model
 \cite{witten,mohapatrae6}. 
One has to rely on 
 symmetries that may arise in superstring models on
specific Calabi-Yau space to 
conveniently 
restrict the Yukawa couplings.
If we have $\mu\equiv 0$ in matrix \ref{matrix2}, this leads to a 
massless neutrino and a massive Dirac neutrino. 
There would be
neutrino mixing even if the light neutrino remains strictly
massless.
If we
include a possible Majorana mass term for the 
$S$-fermion of order
$\mu\simeq 100$ GeV we get similar values of the 
neutrino mass as in the previous case.

It is worthy to mention that mass matrices as that one 
appearing in expression~\ref{matrix2} have 
been proposed  without embedding in a supersymmetric 
 or any other deeper theoretical  
 framework. In this case  small tree level neutrino masses are obtained
 without making use of large scales.
For example, the model proposed by Ref.\cite{caldwell1} (see also Ref.\cite{valle1})
which incorporates by hand 
additional iso-singlet neutral fermions. 
The smallness of neutrino masses is explained directly from the, otherwise
 left unexplained, smallness of the
parameter  $\mu$ in such a model.

\subsection{Neutrino masses and extra dimensions}

Recently,  models where space-time is endowed  with extra dimensions 
(4+$n$) have received some interest~\cite{dvali}. 
It has been realized that the fundamental scale of 
gravity need not be  the 4-dimensional 
``effective'' Planck scale $M_P$ 
   but a new scale $M_f$, as low as $M_f\sim$ TeV. 
The observed Planck scale $M_P$ is 
then related to 
$M_f$ in $4+n$ dimensions, 
by 
$$\eta^2\equiv \left (\frac{M_f}{M_P}\right )^2 \sim \frac{1}{ M_f^n R^n}$$ 
where $R$ is the typical length  of the extra dimensions. 
\ie, the coupling is $M_f/M_P \simeq 10^{-16}$ for $M_f \simeq 1$
TeV. 
For $n=2$, the radii  $R$ of the extra
dimensions are of the order of the 
millimeter, which could be hidden from
many, extremely precise, measurements that exist 
 at present but it would  give hope to probe the
concept of hidden space dimensions (and gravity itself) 
by experiment in the near future.

According to current theoretical frameworks
(see for example Ref.~\cite{dvali}), 
all the SM  group-charged particles are 
 localized on a $3$-dimensional 
hyper-surface `brane' embedded in the bulk of the 
$n$  extra dimensions. 
 All the particles split in two
categories, those that live on the brane and those which exist
every where, as `bulk modes'. 
In general,  any coupling between the brane and 
the bulk modes are suppressed by the
 geometrical  factor $\eta$.
Graviton and possible other neutral states  belongs to the second category.
The observed weakness of gravity can be then interpreted as a
 result of the  new space dimensions in which 
gravity can  propagate.

The small coupling above can  also be used to explain 
the smallness of the neutrino mass~\cite{smirnov}. 
The left handed neutrino $\nu_L$
having weak isospin and hypercharge must reside on the brane. 
Thus it can get a naturally small Dirac mass through 
the mixing with some bulk fermion which can be 
interpreted as right handed
neutrinos $\nu_R$: 
$$L_{mass,Dirac}\sim h \eta H \bar{\nu}_L \nu_R.$$ 
Here $ H,h$ are the Higgs doublet fields and a Yukawa coupling.
After EW breaking this interaction will generate
the Dirac mass
$m_D = h v \eta \simeq 10^{-5}\  \mathrm{eV}$.
The right handed neutrino $\nu_R$ has a whole tower of Kaluza-Klein 
 relatives $\nu_{iR}$.
The masses of these states are given by
$m_{i} =  i/R $, the $\nu_L$ couples with all 
with the same mixing mass. 
We can write the mass Lagrangian as  
$L=\bar{\nu}_L M \nu_R$ 
where $\nu_L=(\nu_L,\tilde{\nu}_{1L}, ...)$ ,
 $\nu_R=(\nu_0R,\tilde{\nu}_{1R}, ...)$ 
and the resulting mass matrix $M$ being:
\begin{equation}
M = \pmatrix{
 m_D & \sqrt{2}  m_D  & \sqrt{2}  m_D & .&\sqrt{2}  m_D &. \cr
        0 & 1/R & 0 & .&0 & .\cr
        0 & 0 & 2/R & . & 0&.\cr
        . &.  &  .  & . & k/R&.\cr
        . &.  &  .  & . &  . &. }
\label{extra2}
\end{equation}
The eigenvalues of the matrix $M M^{\dag}$ are given by a 
transcendental equation. In the limit, 
$m_D R \rightarrow 0$, or  $m_D\rightarrow 0$, the eigenvalues 
are $\sim k/R$, $k\in Z$ with a doubly-degenerated zero eigenvalue.

Other examples can be  considered which incorporates a LR 
symmetry 
(see for example Ref.~\cite{mohapatra2}),
 a $SU(2)_R$ right handed neutrino is
assumed to live on the brane together with the standard one. 
In this class of models, it has been 
shown that the left handed neutrino is exactly massless whereas
assumed bulk sterile neutrinos have masses related to the size of the
extra dimensions. They are  of order $10^{-3}$ eV, 
if there is at least one extra dimension with size 
in the micrometer range.

\subsection{Family symmetries and neutrino masses}

The observed mass and mixing interfamily  hierarchy 
in the quark and, presumably in the lepton sector might 
 be a    consequence of the existence of  
a number of  $U(1)_F$ family symmetries~\cite{froggatt}. 
The observed intrafamily hierarchy, 
the fact that for each family  
$m_{up}>> m_{down}$, seem to require one 
of these to be anomalous \cite{familons1,familons2}.

A simple model with one family-dependent anomalous $U(1)$ 
 beyond the SM was first proposed in Ref.\cite{familons1} to 
produce the observed Yukawa hierarchies, the anomalies 
 being canceled by the Green-Schwartz mechanism which as 
 a by-product is able to fix the Weinberg angle (see also Ref.\cite{familons2}).

Recent developments includes the model proposed in Ref.\cite{RAMOND},
which  is inspired by models generated by the 
$E_6\times E_8$ heterotic string. 
The gauge structure of the  model 
is that of the SM augmented 
 by three Abelian $U(1)$ symmetries $X,Y^{1,2}$, the first one  
  is anomalous and family independent. 
Two of the them, the non-anomalous ones, have 
specific  dependences on the three chiral families designed 
to reproduce the Yukawa hierarchies. 
There are right handed 
neutrinos which trigger neutrino masses by the see-saw 
mechanism.

The three symmetries  $X,Y^{1,2}$ are spontaneously broken at a  high 
 scale M by stringy effects.
It is assumed that three fields $\theta_i$ acquire a vacuum 
 value. The $\theta_i$ fields are singlets under the SM 
symmetry but not under the $X$ and $Y^{1,2}$ symmetries.
In this way, 
the Yukawa couplings appear as the effective
operators after $U(1)_F$ spontaneous symmetry breaking. 

For neutrinos we have~\cite{ramond3} the mass Lagrangian
\begin{eqnarray} 
L_{mass}&\sim  & h_{ij} L_i H_u N_j^c 
\lambda^{q_i+n_j} +  M_N \xi_{ij} N_i^c N_j^c 
\lambda^{n_i+n_j} 
\nonumber
\end{eqnarray}
where $h_{ij}, \xi_{ij} \simeq
O(1)$. 
The parameter $\lambda$  determine the mass and mixing hierarchy,
$\lambda~=~\langle\theta\rangle/M\sim\sin\theta_c$
 where $\theta_c$ is the Cabibbo angle. 
The $q_i, n_i$ are the $U(1)$ charges 
assigned  respectively to left handed leptons  $L$ 
and right handed neutrinos $N$.

These coupling generate the following 
mass matrices for neutrinos:
\begin{eqnarray}
m_{\nu}^D &= & diag(\lambda^{q_1}, \lambda^{q_2} , \lambda^{q_3})\
\hat{h}\ diag(\lambda^{n_1}, \lambda^{n_2} , \lambda^{n_3})
\langle H_u \rangle ,\nonumber\\ M_{\nu} & = & diag(\lambda^{n_1},
\lambda^{n_2} , \lambda^{n_3})\ \hat{\xi}\ diag(\lambda^{n_1},
\lambda^{n_2} , \lambda^{n_3}) M_N.
\end{eqnarray}
{}From these matrices, the see-saw mechanism gives the formula 
 for light neutrinos: 
$$ m_{\nu} \simeq
\frac{\langle H_u \rangle^2}{M} diag(\lambda^{q_1}, \lambda^{q_2},
\lambda^{q_3})\ \hat{h}\ \hat{\xi}^{-1}\ \hat{h}^{T}\
diag(\lambda^{q_1}, \lambda^{q_2}, \lambda^{q_3}).$$ 
The neutrino mass mixing matrix depends only on the charges 
 assigned to the left handed neutrinos, by a cancellation 
 of right handed neutrino charges by virtue  of the see-saw
 mechanism.
There is freedom in assigning charges $q_i$.
If the
charges of the second and the third generations of leptons are
equal ($q_2 = q_3$), then one is lead to a mass
matrix which have the following structure:
\begin{equation}
m_\nu  \sim   \left(
\begin{array}{ccc}
        \lambda^6 & \lambda^3 & \lambda^3 \\
        \lambda^3 & a & b\\
        \lambda^3 & b & c
        \end{array} \right) .
\end{equation}
where $a, b, c \sim O(1)$. This matrix can be 
 diagonalized by a large
$\nu_2 - \nu_3 $ rotation, it is consistent with a 
 large $\mu-\tau$
mixing. In this theory, explanation of 
the large neutrino mixing is reduced to a
theory of prefactors in front of powers of the parameter 
$\lambda$.

\section{Cosmological Constraints}
\label{sectioncosmos}
\subsection{Cosmological mass limits and  Dark Matter}

There are some indirect constraints on neutrino masses 
provided by cosmology. 
The most relevant is the constraint which follows from
demanding that the energy density in neutrinos should not
be too high. At the end of this section we will deal with 
 some other limits as the lower mass limit obtained from 
galactic phase space requirements or limits on the 
 abundance of additional weakly interacting light particles.

Stable neutrinos with low masses ($m_\nu\lsim\ 1$ MeV)
make a contribution to the total energy density of the universe
which is given by:
\begin{eqnarray}
\rho_\nu&=& m_{tot}\ n_\nu
\end{eqnarray}
where the total mass $m_{tot}=\sum_\nu (g_\nu/2) m_\nu$, with 
the number of  degrees of freedom
$g_\nu=4(2)$ for Dirac (Majorana) neutrinos.
The number density of 
the neutrino sea is related to that one of 
 photons by entropy conservation in the adiabatic expansion
 of the universe,  $n_\nu=3/11\ n_\gamma$, and this last 
 one is very accurately obtained from the CMBR measurements, 
 $n_\gamma= 410.5$ cm$^{-3}$ (for a Planck spectrum with 
 $T_0=2.725\pm 0.001\ K\ \simeq 2.35 \times 10^{-4}$ eV).
Writing $\Omega_\nu=\rho_\nu/\rho_c$, where $\rho_c$ is the
 critical energy density of the universe
($\rho_c=3 H_{0}^2/8 \pi G_N$), we have
($m_\nu>> T_0$)
\begin{eqnarray}
\Omega_\nu h^2& =& 10^{-2}\ m_{tot}\  ( \mathrm{eV}),
\label{cosmoneutrino}
\end{eqnarray}
where $h$ is the reduced Hubble constant,
recent analysis 
\cite{hubble} give  the favored  value: 
$h=0.71\pm0.08$.

Constrained by requirements from BBN Nucleosynthesis, 
 galactic structure formation and large scale observations,
increasing evidence (luminosity-density relations, galactic rotation 
 curves,large scale flows) suggests that
\cite{darkmatter}
\begin{equation} 
\Omega_{M}  h^2= 0.05 - 0.2,
\end{equation}
where $\Omega_{M}$ is the total mass density of the universe, 
as a fraction of the critical density $\rho_c$.
This $\Omega_{M}$ includes contributions from a 
 variety of sources: photons, baryons,  non-baryonic 
Cold  Dark Matter (CDM) and Hot Dark Matter (HDM).

The two first components are rather well known.
 The photon density is very well known to be quite small:
$ \Omega_\gamma h^2= 2.471\times 10^{-5}$.
The deuterium
abundance BBN constraints~\cite{deuterium} on the baryonic matter density 
($\Omega_B$) of
the universe $0.017 \leq \Omega_B h^2 \leq 0.021.$

The hot component, 
HDM is constituted by  relativistic  
long-lived particles with masses  
much less than $\sim 1$ keV, in this category would enter 
the neutrinos.  
Detailed simulations of structure formation fit the observations only when
one has some 20 \% of HDM (plus $80\%$ CDM), 
the best fit being two neutrinos with a
total mass of 4.7 eV. 
There seems to be however some kind of conflict within cosmology
itself: observations of distant objects favor a large cosmological
constant instead of HDM (see Ref.\cite{HDM} and references
therein). One may conclude that the HDM
part of $\Omega_M$ does not exceed 0.2. 

Requiring that $\Omega_\nu<\Omega_M$, we obtain
 $\Omega_{\nu} h^2 \lsim\  0.1$. From here and from Eq.\ref{cosmoneutrino},  
we obtain the cosmological upper bound 
on the neutrino mass 
$$m_{tot}\lsim\ 8\ \mathrm{eV}.$$

Mass limits, in this case lower limits,  for heavy neutrinos ($\sim 1$ GeV) can also be obtained
along the same lines. 
The situation gets very different if the neutrinos are unstable, one 
gets then joint bounds on mass and lifetime, then mass limits above 
 can be avoided. 

There is a limit  to the density of neutrinos (or weak 
 interacting dark matter in general) which  can be accumulated in the
halos of astronomical objects (the {\em Tremaine-Gunn} limit): 
if neutrinos form part of the galactic bulges phase-space restrictions from the 
 Fermi-Dirac distribution  implies a lower limit  on the 
neutrino mass \cite{peacock}:
$$m_\nu\gsim\  33 \ eV.$$

The abundance  of additional weakly interacting light particles, such as a
light sterile $\nu_s$, is constrained by BBN since it would enter into
equilibrium with the active neutrinos via neutrino oscillations.
A limit on the mass differences and mixing angle with another 
active neutrino  of the type 
$\Delta m^2 \sin^2 2\theta\lsim 3\times 10^{-6}$ eV$^2$ should be
fulfilled in principle.
{}From here is deduced that the effective number of neutrino 
species is 
$$N_\nu^{eff}< 3.5-4.5.$$ 
However systematical uncertainties in the
derivation of the BBN bound make it too unreliable to be
taken at face value and can eventually be avoided \cite{foot}.

\subsection{ Neutrino masses and lepton asymmetry}

In supersymmetric LR symmetric models, inflation, 
baryogenesis (or
leptogenesis) and neutrino oscillations can  become closely linked. 

Baryosinthesis in  GUT theories is in general inconsistent with 
an inflationary universe. The exponential expansion during  
 inflation will wash out any  baryon asymmetry generated previously 
at GUT scale. One way out of this difficulty is to generate the 
baryon or lepton asymmetry during the process of reheating at the 
 end of the inflation. In this case the physics of the scalar field
that drives the inflation, the inflaton, would have to violate CP 
(see Ref.\cite{peacock} and references therein).

The challenge of any baryosinthesis  model is to predict the
 observed asymmetry which 
 is usually written as a baryon  to photon (number or 
 entropy) ratio. 
The baryon asymmetry is defined as 
\begin{eqnarray}
n_B/s\equiv \left (  n_b-n_{\overline{b}}\right )/s.
\end{eqnarray}
At present there is only matter and not known antimatter, $n_{\overline{b}}\sim 0$.
The entropy density $s$ is completely dominated by the 
contribution of relativistic particles so is proportional to 
 the photon number density which is very well known from CMBR 
 measurements, at present $s=7.05\ n_\gamma$.
Thus, 
$n_B/s\propto n_b/n_\gamma$. From BBN we know that 
$n_b/n_\gamma=(5.1\pm 0.3)\times 10^{-10}$ so we arrive to 
$n_B/s=(7.2\pm 0.4)\times 10^{-11}$ and from here we obtain equally the 
 lepton asymmetry ratio.

It was shown in Ref.~\cite{khalil} that  hybrid inflation can be
successfully realized in a SUSY LR symmetric model  with  gauge group
$G_{PS}=SU(4)_c\times SU(2)_L \times SU(2)_R$. The inflaton sector  of
this model consists of the two complex scalar fields $S$ and
$\theta$  which 
at the end of inflation  oscillate about the SUSY minimum and respectively decay
into a pair of right-handed sneutrinos ($\nu^c_i$) and neutrinos.
In this model, a primordial lepton asymmetry is
generated~\cite{yanagita} by the decay of the superfield $\nu^c_2$ which
emerges as the decay product of the inflaton. 
The superfield $\nu^c_2$ decays
into electroweak Higgs and (anti)lepton superfields. 
This lepton
asymmetry is subsequently partially converted into baryon asymmetry by
 non-perturbative EW sphalerons. 

The resulting lepton asymmetry~\cite{Laz3}  can be written as a 
function of a number of parameters among them the neutrino masses 
and mixing angles and compared with the observational constraints above.

It is highly non-trivial
that solutions satisfying the constraints above and 
other physical  requirements can be found with
natural values of the model parameters.
In particular, it is shown that  the values of the neutrino masses and 
 mixing angles which predict sensible values for the baryon or lepton 
asymmetry 
turn out to be also 
consistent with values required to solve the solar neutrino problem.

\section{Phenomenology of Neutrino Oscillations}
\subsection{Neutrino Oscillation in Vacuum}
If the neutrinos have nonzero mass, 
by the basic postulates of the quantum theory
there will be in general
mixing among them  as in the case of quarks. 
This mixing will be observable at macroscopic distances from 
the production point and therefore will have practical 
 consequences only if the {\em difference}  of masses of the 
 different neutrinos is very small, 
typically $\Delta m\lsim 1$ eV.

In presence of masses,  
weak ($\nu_w$) and mass ($\nu_m$)
 basis of eigenstates are differentiated. 
To transform between them
 we need an unitary matrix $U$.
Neutrinos can only be created and detected as a result 
 of  weak processes, at origin we have  a weak eigenstate: 
$$\nu_w(0)= U \nu_m(0).$$
We can easily construct an heuristic theory of 
 neutrino oscillations if we ignore spin effects as follows. 
After a certain time the system has evolved into
$$\nu_m(t)=\exp (-i H t) \nu_m(0)$$ 
where $H$ is the Hamiltonian of the system, free evolution in 
 vacuum is characterized by $H=diag(\dots E_i \dots)$ where 
$E_i^2=p^2+m_i^2$. In most cases of interest ($E\sim$MeV, $m\sim$eV), 
it is 
 appropriated the ultrarelativistic limit: in this limit $p\simeq E$ and 
$E\simeq p+m^2/2p$. The effective neutrino 
Hamiltonian can then be written
 $H^{eff}=diag(\dots m_i^2\dots)/2E $ and 
$$\nu_w(t)= U \exp (-i H^{eff} t) U^\dagger \nu_w(0)=\exp(-i H^{eff}_w t) \nu_w(0).$$
In the last expression we have written the effective Hamiltonians in the 
weak basis 
$H^{eff}_w~\equiv~M^2/2E$ with 
$M\equiv U\ diag(\dots m_i^2 \dots) U^\dagger$.
This derivation can be put in a firm basis and 
one finds again the same expressions as 
 the first terms of rigorous expansions in $E$, 
see for example the treatment 
 using Foldy-Woythusen transformations  in Ref.\cite{BILENKY}.

The
 results of the neutrino oscillation experiments are usually
analyzed under the simplest assumption of oscillations between two
neutrino types, in this case the mixing matrix $U$ is the well 
 known 2-dimension orthogonal rotation matrix depending on a
 single parameter $\theta$.
If we repeat all the computation above for this particular case, we find 
for example 
that  the probability that a weak interaction
eigenstate neutrino $(\nu_e)$ has oscillated to other weak
interaction eigenstate neutrino $(\nu_{\mu})$ after traversing a
distance $l(= c t)$ is
\begin{equation}
P(\nu_e \rightarrow \nu_{\mu}; l)= 
\sin^2 2\theta \sin^2 
\left (\frac{l}{l_{osc}} \right ) 
\label{oscil1}
\end{equation}
where the oscillation length is defined by 
$1/l_{osc}\equiv \delta m^2 l/4 E$  
and $\delta m^2 = m_1^2-m_2^2$. 
Numerically, in practical units, it turns out that 
$$\frac{\delta m^2l}{4 E} \simeq 1.27\ \frac{\delta m^2 (eV^2)\  l(m) }{E(MeV)}.$$
These probabilities depend on two factors: a
mixing angle factor $\sin^2 2 \theta $ and a kinematical factor
which depends on the distance traveled, on the momentum of the
neutrinos, as well as on the difference in the squared mass of the
two neutrinos. Both, the mixing factor $\sin^2 2\theta $ and the 
 kinematical factor should be of
$O(1)$ to have a significant oscillations. 

\subsection{Neutrino Oscillations in Matter}

When neutrinos propagate in matter, a subtle but potentially very 
important effect, the MSW effect,
takes place which alters the way in which neutrinos oscillate into
one another. 

In matter the neutrino experiences scattering and absorption, this last one
 is always negligible.
 At very low energies, coherent elastic forward scattering is the 
 most important process. As in optics, the net effect is 
the appearance of a phase difference, refractive index or equivalently 
 a neutrino effective mass.

This effective mass can considerable change depending on the 
 densities and composition of the medium, it  depends 
also on the nature of the neutrino.
In the neutrino case 
 the medium is  flavor-dispersive: the matter 
is usually nonsymmetric with respect $e$ and $\mu,\tau$ and the 
effective mass is different for the different 
 weak eigenstates  \cite{msw2}.

This is explained as follows for the simpler and most important case, the 
solar electron plasma.
The electrons in the solar medium have charged current
interactions with $\nu_e$ but not with $\nu_{\mu}$ or $\nu_{\tau}$.
The resulting interaction energy is given by $ H_{int} = \sqrt{2}
G_F N_e$, where $G_F$ and $N_e$ are the Fermi coupling and the
electron density. The corresponding neutral
current interactions are identical for all neutrino species and
hence have no net effect on their propagation.
Hypothetical sterile neutrinos would have no interaction at all.
 The effective global Hamiltonian in flavor space is now the sum of two terms, the 
 vacuum part we have seen previously and the new interaction energy:
$$H^{eff,mat}_w=H^{eff,vac}_w+ H_{int}
\pmatrix{
1 & 0 & 0\cr 
0 & 0 & 0\cr 
0 & 0 & 0} 
.$$

The practical consequence  of this effect is that the oscillation 
 probabilities of the neutrino in matter could largely 
increase due to resonance phenomena \cite{msw1}.
In matter, for the two dimensional case and in analogy with 
vacuum oscillation, one  defines an effective mixing angle  as
\begin{equation}
\sin 2\theta_M = \frac{ \sin2\theta/l_{osc}}
{\left[(\cos 2\theta/l_{osc} - G_F N_e/\sqrt{2})^2
+ ( \sin 2\theta/l_{osc} )^2 \right]^{1/2}}.
\end{equation}
The presence of the term proportional to the electron density
can give  rise to  a resonance. There is a critical
density 
$N_e^{crit}$, 
given by 
$$ N_e^{crit} = \frac{\delta m^2\cos 2\theta}{2\sqrt{2} E G_F},$$ 
for which the matter mixing
angle $\theta_M$ becomes maximal $(\sin 2 \theta_M \rightarrow
1)$, irrespective of
 the value of  mixing angle $\theta$. The
probability that $\nu_e$ oscillates
 into a $\nu_{\mu}$ after
traversing a distance $l$ in this medium is given by
Eq.(\ref{oscil1}), with two differences. First $\sin 2\theta
\rightarrow \sin 2\theta_M$. Second, the kinematical factor differ
by the replacement of $\delta m^2 \rightarrow \delta m^2 \sin
2\theta$. Hence it follows that, at the critical density,
\begin{equation}
P_{\mathrm{matter}}(\nu_e \rightarrow \nu_{\mu}; l)_
{(N_e =N_e^{\mathrm{crit}})} = \sin^2 \left (  \sin2\theta\ 
\frac{l}{l_{osc}}\right ) . 
\label{oscil2}
\end{equation}
This formula shows that one can get full conversion of a $\nu_e$
weak interaction eigenstate into a $\nu_{\mu}$ weak interaction
eigenstate, provided that the length $l$ and the energy $E$
satisfy the relations 
$$ \sin 2\theta\
\frac{l}{l_{osc}} = \frac{n \pi}{2} ; \quad n=1,2,..$$ 
There is a second interesting
limit to consider. This is when the electron density $N_e$ is so
large such that $\sin 2 \theta_M \rightarrow 0$
 or $\theta_M \rightarrow \pi/2$. In this
limit, there are no oscillations in matter because 
$\sin2\theta_M$ vanishes and we have
$$
P_{\mathrm{matter}}(\nu_e \rightarrow
\nu_{\mu}; l)_{\left (N_e \gg \frac{\delta m^2}{2 \sqrt{2} E G_F}\right )}
\rightarrow 0.
$$

\section{Experimental evidence and phenomenological 
 analysis}

In the second part of this review, we will consider the 
existing experimental situation. It is fair to say that at present 
there are 
at least an equal number of positive as negative (or better ''non-positive'')
 indications in favor of neutrino masses and oscillations.

\subsection{Laboratory, reactor and accelerator results.}

No indications in favor of a 
non-zero neutrino masses have been found in direct
kinematical searches for a neutrino mass.

{}From the measurement of the high energy part of the
tritium $\beta$ decay  spectrum, upper limits on the electron neutrino mass 
are obtained.
The two more sensitive experiments in this field, Troitsk  \cite{troitsk} and Mainz \cite{mainz}, obtain 
 results which are plagued  by
interpretation problems: apparition of negative
 mass squared and bumps at the end of the spectrum.

In the Troistk experiment, 
the shape of the observed spectrum proves to be in accordance with
classical shape besides a region $\sim$ 15 eV below the end-point, 
where a small bump is observed;
there are indications   of a periodic 
shift of the position of this bump with a period of ``exactly'' $0.504\pm 0.003$ year \cite{troitsknew}.
After accounting for the bump, they derive the limit
$m_{\nu e}^2=-1.0\pm 3.0\pm 2.1$ eV$^2$, or 
 $m_{\nu e}< 2.5$ eV (95
\cite{troitsknew}.

The latest published results by the Mainz group
 leads to 
$m_{\nu e}^2=-0.1\pm 3.8\pm 1.8$ eV$^2$ (1998 ``Mainz data 1''),
{}From which an upper limit of  $m_{\nu e}< 2.9$ eV \cite{mainz}
(95\% C.L., unified approach) is obtained.
Preliminary data (1998 and 1999 measurements) provide a limit  
 $m_{\nu e}< 2.3$ eV \cite{mainznew}.
Some
indication for the anomaly, reported by the Troitsk group, was found, but
its postulated half year period is not supported by their data.

Diverse  exotic explanations 
 have been proposed to explain the Troitsk bump
 and their seasonal dependence.
The main feature of the effect might
 be ``phenomenologically'' interpreted, not without problems, 
as $^3$He capture of relic neutrinos present in a high 
density cloud around the Sun \cite{troitsk,stephe}.

The Mainz and Troitsk ultimate sensitivity  expected to be
limited by systematics lies at the $\sim 2 $ eV level.
In the near future, it is planned a new large tritium $\beta$ 
experiment with sensitivity $0.6- 1$ eV \cite{mainznew}.

Regarding the heavier neutrinos, other kinematical limits are the following:

\begin{itemize}

\item[a)] Limits for the muon neutrino mass have been derived using the decay
channel $\pi^+\to\mu^+ \nu_\mu$ at intermediate energy accelerators (PSI, LANL).
The present limits are $m_{\nu \mu}\lsim 160 $ keV \cite{PSI}.

\item[b)]
A tau neutrino mass of less than 30 MeV is well established and confirmed by several experiments:
limits of 28, 30 and 31 MeV have also been  obtained by the OPAL, CLEO
 and ARGUS experiments respectively 
(see Ref.\cite{OPAL} and references therein).
The best upper limit for the $\tau$ neutrino mass has been derived using the
decay mode $\tau\to 5 \pi^\pm \nu_\tau$ by the ALEPH collaboration
\cite{ALEPH}:
$m_{\nu\tau}<$ 18 MeV (95\% CL).
\end{itemize}

Many experiments on the search for neutrinoless double-beta decay
[$(\beta\beta)_{0\nu}$],
$$(A,Z)\to\ (A,Z+2)\to 2\ e^-,$$
have been performed. This process is possible only
if neutrinos are massive and Majorana particles. The matrix element of the process
is proportional to the effective Majorana mass
$\langle m\rangle=\sum \eta_i U_{ei}^2 m_i$.
Uncertainties in the precise value of upper limits are 
relatively large   since
they  depend on  theoretical calculations of   nuclear matrix elements.
{}From the non-observation of
$(\beta\beta)_{0\nu}$,
the Heidelberg-Moscow experiment gives the most stringent limit on the
Majorana neutrino mass. 
After 24 kg/year of data 
 \cite{heidelberg99} (see also earlier
 results in Ref.\cite{heidelberg}), 
they set a lower limit on the
half-life of the neutrinoless double beta decay in $^{76}$Ge of
$T_{1/2}>5.7\times 10^{25}$ yr at 90\% CL, thus excluding an
effective Majorana neutrino mass
$\mid\langle m\rangle\mid >0.2 $ eV (90\% CL). This result 
allows
to set strong constraints on degenerate neutrino mass models.
In the next years it is expected an increase in  sensitivity  allowing
limits down  to the
$\mid\langle m\rangle\mid \sim 0.02-0.006$ eV levels (GENIUS I and II 
 experiments, \cite{genius}).

Many short-baseline (SBL)
neutrino oscillation experiments with reactor and accelerator neutrinos did not find any evidence of neutrino oscillations.
For example experiments looking for be
$\overline{\nu}_e\to\overline{\nu}_e$ or
${\nu}_\mu\to{\nu}_\mu$
dissaperance (Bugey, CCFR \cite{Bugey,CCFR}) or oscillations
$\overline{\nu}_\mu\to\overline{\nu}_e$ (CCFR,E776\cite{CCFR,E776}).

\begin{figure}[ht]
\centering
\begin{tabular}{cc}
\mbox{\psfig{file=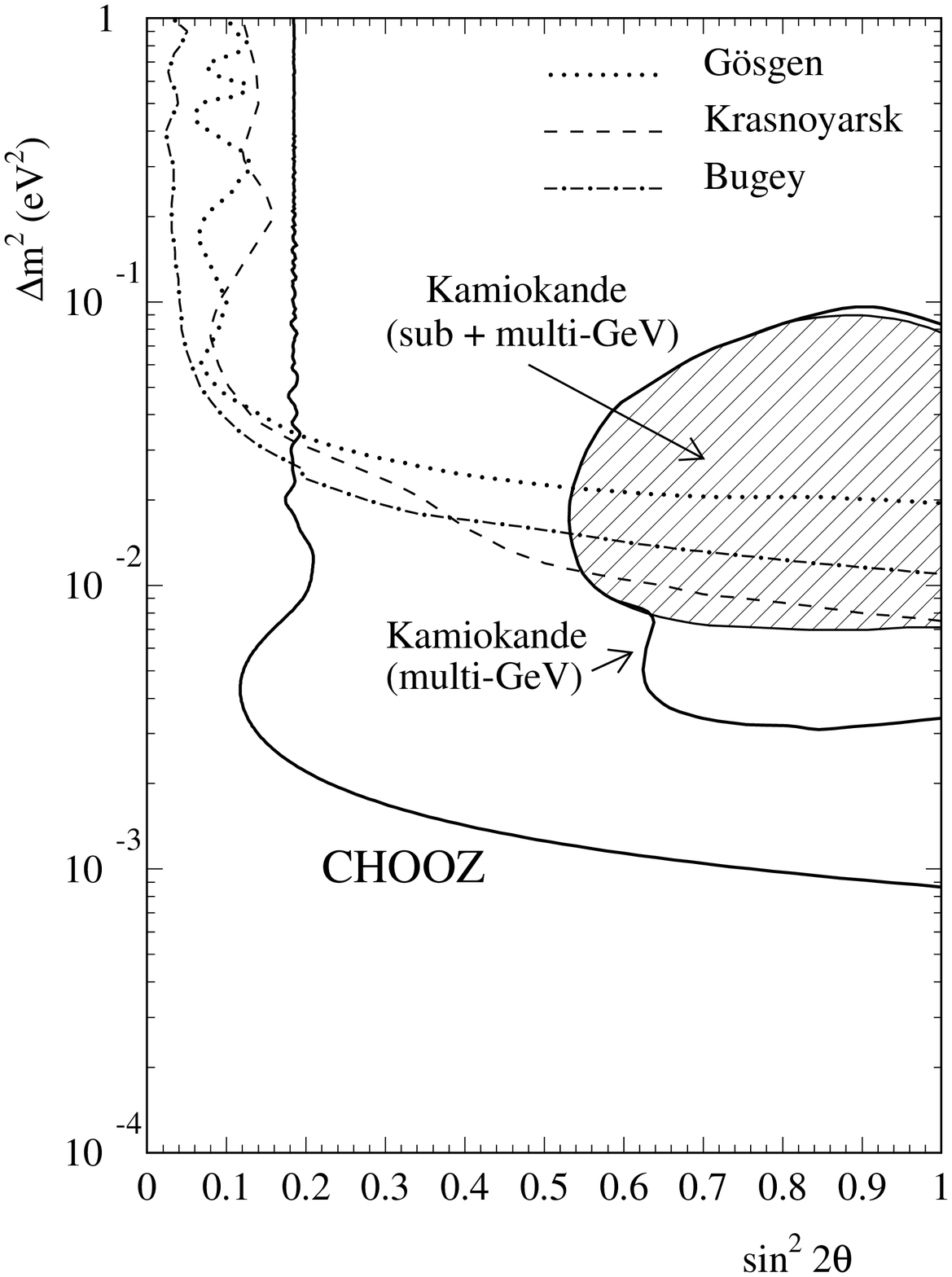,height=7cm,width=6cm}}&
\epsfig{file=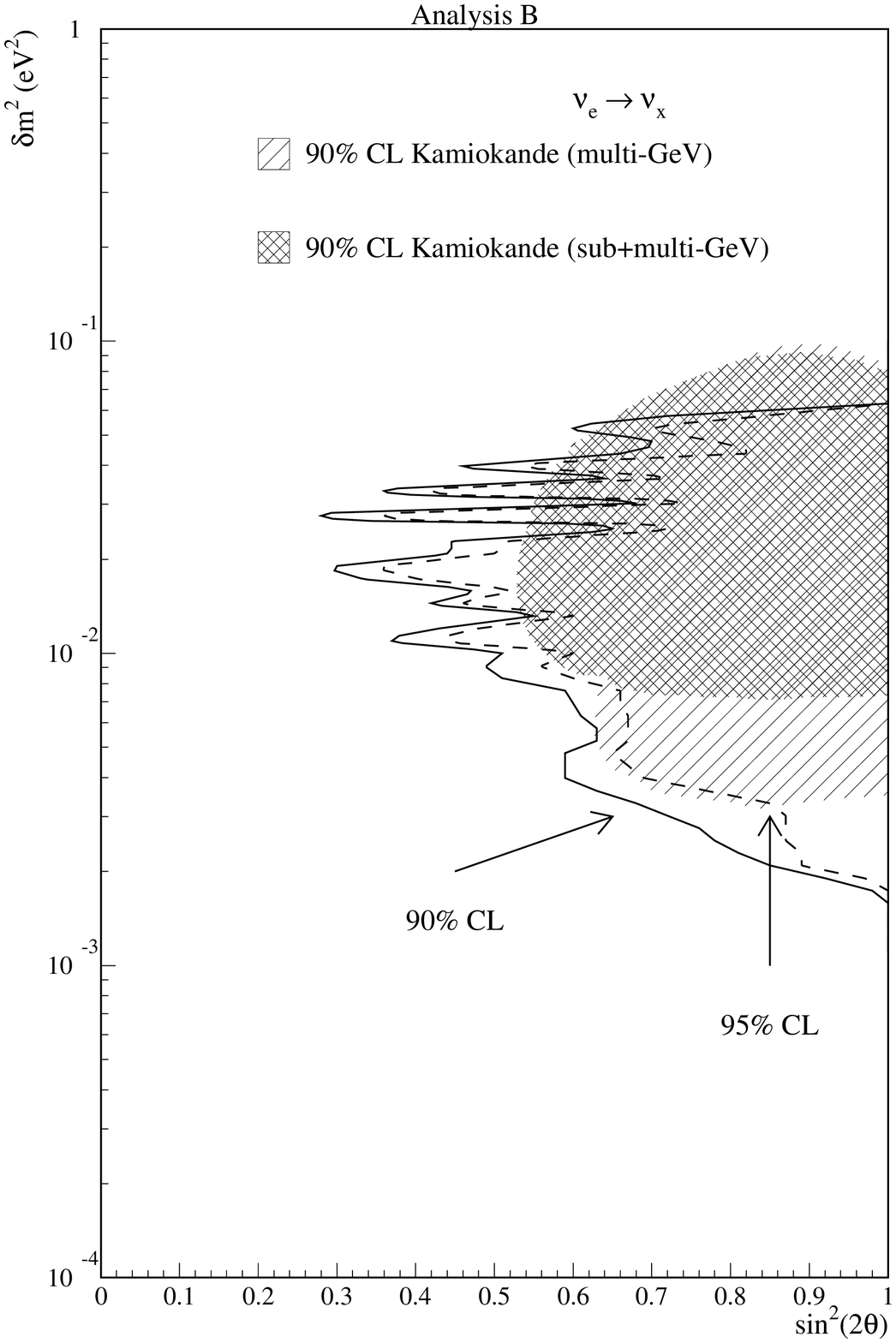,height=7cm,width=6cm}
\end{tabular}
\vspace{0cm} \caption{(Left) The 90\% C.L. exclusion plot for
CHOOZ, compared with previous experimental limits and with the
KAMIOKANDE allowed region. (From Ref.\protect\cite{CHOOZ}).
(Right) Exclusion plot contours obtained from the ratios of the
positron spectra from the two reactors (From
Ref.\protect\cite{CHOOZ99}). } \label{fCHOOZ}
\end{figure}

The first reactor long-baseline (L$\sim$ 998-1115 m) 
neutrino oscillation experiment
CHOOZ found  no evidence for neutrino oscillations in the $\nuebar$
disappearance mode \cite{CHOOZ,CHOOZ99}.
CHOOZ results are important for the atmospheric deficit problem:
 as  is
seen in Fig.(\ref{fCHOOZ}) they are incompatible with
an  $\nue\to \numu$ oscillation hypothesis
for the solution of the
atmospheric problem.
Their latest results \cite{CHOOZ99}  imply an exclusion region
in the plane of the two-generation mixing parameters
(with normal or sterile  neutrinos)
given
approximately by $\Delta m^2 > 0.7~10^{-4}\units{eV^2}$ for maximum mixing and
$\sinsq > 0.10$ for large $\dmsq$ (as shown approximately in Fig.(\ref{fCHOOZ}) (left) which corresponds to early results).
 Lower sensitivity results, based only on the comparison of the
positron spectra from the two different-distance nuclear reactors, has also
been presented, they are shown in Fig.(\ref{fCHOOZ}) (right). These  are independent of the absolute normalization of
the antineutrino flux, the cross section and the target and detector
characteristics and are able alone to almost completely exclude the
SK allowed oscillation region \cite{CHOOZ99}.


The Palo Verde Neutrino Detector 
 searches for neutrino oscillations via the disappearance
of electron anti-neutrinos produced by a nuclear reactor at a 
distance $L\sim 750-890$ m.
The experiment has been taking neutrino data 
since October 1998 and will continue taking data until the end of 2000 reaching its ultimate sensitivity. 
The analysis of the  1998-1999 data (first 147 days of operation) 
\cite{paloverdenew} yielded no evidence for the
existence of neutrino oscillations.
The ratio of observed to expected number of events:
$$\frac{\overline{\nu}_{e,obs}}{\overline{\nu}_{e,MC}}=1.04\pm 0.03\pm0.08.$$
 The resulting $\overline{\nu}_e\to\overline{\nu}_x$ exclusion plot is very similar to the CHOOZ one. Together with results from CHOOZ and SK, concludes that the atmospheric neutrino anomaly is very unlikely to be due to
 $\overline{\nu}_\mu\to\overline{\nu}_e$ oscillation.


Los Alamos LSND experiment  has reported
indications of  possible
$\overline{\nu}_\mu\to\overline{\nu}_e$ oscillations \cite{LSND}.
They search for
$\overline{\nu}_e$'s in excess  of  the number expected from conventional
sources at a liquid scintillator detector located 30 m from a proton
  beam dump at LAMPF. It has been claimed that a 
 $\overline{\nu}_e$ signal has been detected via the reaction
$\overline{\nu}_e p\to e^+ n$ with $e^+$ energy between 36 and 60 MeV,
followed by a $\gamma$ from $n p\to d\gamma$ (2.2 MeV).

The LSND experiment took its last beam on December, 1998.
The analysis
of the complete 1993-1998 data set (see Refs.\cite{yellin,Mills,LSNDnew})
yields  a fitted-estimated excess of $\overline{\nu}_e$ 
 of  $90.9\pm 26.1$.
If this excess is attributed to neutrino oscillations of the 
type $\overline{\nu}_\mu\to\overline{\nu}_e$,
it corresponds to an oscillation probability of
$3.3\pm 0.09\pm 0.05\times 10^{-3}$.
The results of a similar search for
$\nu_\mu\to \nu_e$
oscillations where the (high energy, $60 <E_\nu< 200$ MeV) $\nu_e$ are detected via the CC reaction
$C(\nu_e,e^-) X$
provide a value for the corresponding oscillation probability of
$2.6\pm 1.0\pm 0.5\times 10^{-3}$ (1993-1997 data).

There are other exotic physics explanations of the observed
 antineutrino excess. One example is the lepton-number 
violating decay $\mu^+\to e^+ \overline{\nu}_e \nu_\mu$,
which can explain these observations with a branching ratio
$Br\sim 0.3 \%$, a value which is lower but not very far from
 the respective existing upper limits ($Br< 0.2-1\%$, \cite{PDG98}).


The surprisingly positive LSND result has not been confirmed by the 
KARMEN experiment (Rutherford- Karlsruhe Laboratories).
This experiment, following a similar
experimental setup as  LSND,
searches for $\bar\nu_e$ produced by
$\bar\nu_\mu\to\bar\nu_e$ oscillations at a mean distance of 17.6 m.
The time structure of the neutrino beam
is important for the identification of the neutrino induced reactions
and for the suppression of the cosmic ray background.
Systematic time anomalies not completely understood has been reported which rest credibility to any further
KARMEN claim.
They see an excess of events above the typical
muon decay curve, which is $4.3$ sigmas off 
(1990-1999 data, see Ref.\cite{karmennew}) and which could represent an unknown
instrumental effect.

Exotic explanations as the existence of a
 weakly interacting particle ``X'', for example
 a mixing of active and sterile neutrinos, of a
mass $m_X=m_\pi-m_\mu\simeq$ 33.9 MeV have been proposed as an
alternative solution to these anomalies and their consequences 
extensively 
studied \cite{karmennew,relativekarmenanomaly}. This particle might be
produced in reactions the  $\pi^+\to \mu^+ + X$ and decay as 
 $X\to e^+ e^-\nu$.
KARMEN set upper limits on the visible branching ratio 
$ \Gamma_X=\Gamma ( \pi^+\to \mu^+ + X)/\Gamma ( \pi^+\to \mu^+ + \nu_\mu)
and$  
lifetime $\tau_x$.
{}From their results \cite{karmennew}
one obtains the relation ($1<<\tau_x(\mu\ s)<\sim 10^8$)
$$\frac{\Gamma_X}{\tau_X (\mu\ s)}\sim 10^{-18}.$$

More concretely, the results are as it follows.
About antineutrino signal, the 1990-1995 and early  1997-1998 KARMEN
   data showed inconclusive results:
They found  no events, with an expected background
of $ 2.88 \pm 0.13 $ events, for
$\overline{\nu}_\mu\to\overline{\nu}_e$ oscillations
\cite{KARMEN}.
The results of the search Feb. 1997- Dec. 1999
which include a 40-fold improvement in suppression of
cosmic induced background has
been presented in a preliminary way
\cite{karmennew,steidl99}.
 They find this time  9.5 oscillation candidates
in agreement with the, claimed,
well known background expectation of $10.6\pm 0.6$
events. An upper limit  for the mixing angle is deduced:
$\sin^ 2 2\theta< 1.7\ 10^{-3}$ (90\% C.I.) for large $\Delta m^2\ (= 100$
eV$^2$).
The positive LSND result in this channel could not be completely 
excluded  but they are able to exclude the entire 
LSND favored regions above
2 eV$^2$ and most of the rest of its favored parameter space.

\begin{figure}[ht]
\vspace{0.5cm} 
\centering
\begin{tabular}{cc}
\epsfig{file=karmenfigfeb99b.eps,height=4cm,width=9cm} &
\hspace{-1cm}\epsfig{file=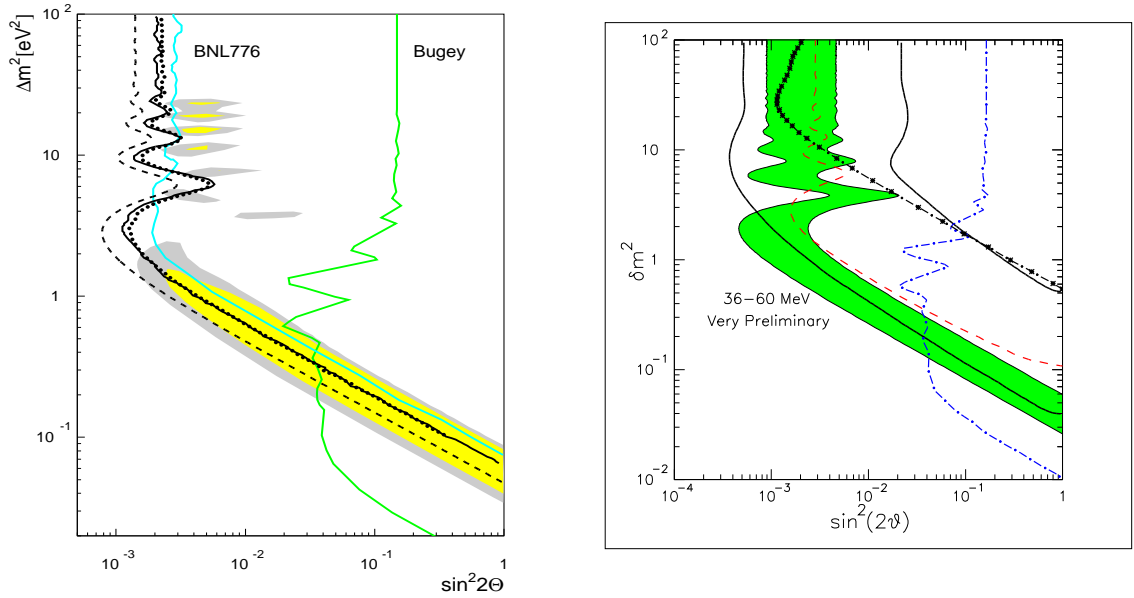,height=7cm,width=7cm}
\end{tabular}
\vspace{0.5cm} \caption{ (Left) KARMEN 2 90\% CL confidence limits
and sensitivity according to the unified approach compared to
LSND,BUGEY and BNL776. The full line is the 90\% C.I. of the
latest available data (feb. 1999) the dotted line the
corresponding sensitivity and the dashed line the expected
sensitivity in year 2001. (From Ref.\protect\cite{steidl99}).
(Right) Comparison of preliminary LSND $\overline{\nu}_\mu\to
\overline{\nu}_e$ results (filled region) with those of KARMEN
(dashed curve), Bugey (dot-dashed) and NOMAD (dot-x-dash).
KARMEN's exclusion curve is 90\% Bayesian confidence
 level, while the LSND region is bounded by
90\% Bayesian upper and lower limits with $36<E_e<60 MeV$. The
pair of smooth curves surrounding the LSND region gives the LSND
90\% C.I. for $\nu_\mu\to\nu_e$ oscillation. (From
Ref.\protect\cite{yellin}).} \label{fLSND}
\end{figure}

In the present phase, the  KARMEN experiment will 
take data until spring 2001.
 At the end of  this period, the
KARMEN sensitivity is expected to be able to
exclude the whole parameter region of evidence suggested by LSND if
no oscillation signal were  found (Fig.\ref{fLSND}).
The first phase of a third pion beam dump
 experiment designed to set the LSND-KARMEN controversy
has been approved to run at Fermilab.
Phase I of ''BooNe'' ( MiniBooNe) expects a 10 $\sigma$ signal
($\sim 1000 $ events)
 and thus will make a decisive statement either proving or
ruling it out. Plans are to run early 2001. Additionally, there is a letter
 of intent of a similar experiment to be carried out at the CERN PS
\cite{BOON,CERNPS}.



  The K2K experiment started in  1999 the era of   
very long-baseline neutrino-oscillation experiment using a 
well-defined neutrino beam.

In the K2K experiment ($L\sim$ 250 km), the neutrino beam generated by the KEK
proton synchrotron accelerator is aimed at the near and far detectors,
which are carefully aligned in a straight line. Then, by comparing the
neutrino events recorded in these detectors, they are able to
 examine the neutrino oscillation phenomenon.  Super-Kamiokande detector itself acts as the far detector.
The K2K near detector complex essentially consists of
 a one kiloton water Cerenkov detector (a miniature
Super-Kamiokande detector).

 A total intensity of $\sim 10^{19}$ protons on target, 
which is about 7\% of the goal of the experiment, 
was accumulated in 39.4 days of data-taking in 1999 \cite{K2K}. 
They obtained 3 neutrino events in the fiducial volume 
of the Super-Kamiokande detector, 
whereas the expectation based on observations in 
the front detectors was $12.3^{ +1.7}_{-1.9}$ neutrino events. 
It corresponds to a ratio of data versus theory $0.84\pm 0.01$. 
Although the preliminary 
results are rather consistent with squared mass difference
 $8\times 10^{-3}$ eV$^2$ and maximal mixing, it is too early to draw any
reliable conclusions about neutrino mixing. 
An complete analysis of oscillation searches from the view points 
of absolute event numbers, distortion of neutrino 
energy spectrum, and  $\nu_e/\nu_\mu$ ratio is still in progress.

\subsection{Solar neutrinos}

Indications in the favor of neutrino oscillations were found in ''all''
solar neutrino experiments 
(along this section and the following ones, we will make 
reference to results appeared in Refs.
\cite{gallex,sage,homestake,sk9812,sk9805,suz1}):
The Homestake Cl radiochemical experiment with sensitivity down
 to the lower energy parts of the $^{8}$B neutrino spectrum and to the
higher $^{7}$Be line \cite{homestake}.
The two radiochemical $^{71}$Ga experiments, SAGE and GALLEX, which are
 sensitive to the low energy pp neutrinos and above \cite{sage,gallex} and the
water Cerenkov experiments Kamiokande and Super-Kamiokande (SK) which
 can observe only the highest energy $^{8}$B neutrinos. Water
 Cerenkov experiments in addition demonstrate directly that the
neutrinos come from the Sun showing that recoil
electrons are scattered in the direction along
the sun-earth axis \cite{sk9812,sk9805,suz1}.

Two important points to remark are:
a) The prediction of the existence of a global neutrino deficit is
hard to modify due to the constraint of the solar luminosity on pp
neutrinos detected at SAGE-GALLEX.
b) The different  experiments are sensitive to neutrinos with different energy ranges and
 combined yield spectroscopic information on the neutrino flux.
Intermediate energy neutrinos arise from intermediate steps
of the thermonuclear solar cycle. It may not be
 impossible to reduce the flux from the last step ($^{8}$B), for example
by  reducing temperature of the center of the Sun, but it seems extremely
 hard to reduce neutrinos from $^7$Be to a large extent, while
keeping a reduction of $^8$B neutrinos production to a modest amount.
If minimal standard electroweak theory is correct, the shape of the
$^8$B neutrino energy spectrum is independent of all solar influences
 to very high accuracy.

\begin{figure}[ht]
\centering
\psfig{file=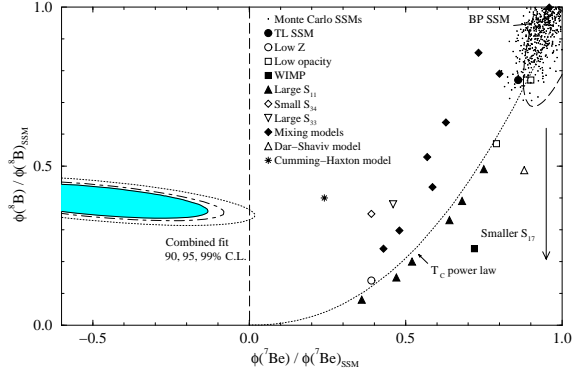,height=10cm} \vspace{-2cm} \caption{ The
severity of the problem with astrophysical solutions. The
constraints on $^8$B and $^7$Be fluxes (considered as free
parameters) from the combined Cl, Ga, and \v{C}erenkov experiments
( 90, 95, and 99\% C.L.)   are shown. The best fit solutions are
obtained for unphysical values. Diverse standard and nonstandard
solar models are shown. [From Hata and Langacker,
Ref.(\protect\cite{hata} and references therein.]} \label{SSM1}
%
\end{figure}

\begin{figure}[ht]
\centering
\psfig{file=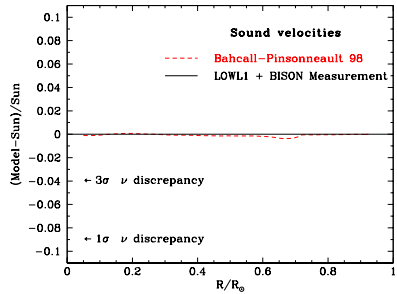,height=9cm} \vspace{-3cm} \caption{ The
excellent agreement between the calculated (solar model BP95
\protect\cite{BP95}) and the measured (Sun) sound speeds. The
fractional error is much smaller than generic fractional changes
in the model, 0.03 to 0.08, that might significantly affect the
solar neutrino predictions. [Adapted from Christensen-Dalsgaard,
Ref.\protect\cite{dalsgard}, as it appears in
\protect\cite{berezinsky}.] } \label{SSM2}
\end{figure}

Unless the experiments are seriously in error, there must be
 some problems with either our understanding of the
Sun  or neutrinos. Clearly, the SSM cannot account for the data
(see Fig.\ref{SSM1})
and possible
highly nonstandard solar models are strongly constrained by
heliosysmology studies [see Fig.(\ref{SSM2})].

There are at least two reasonable 
versions of the neutrino oscillation phenomena which
could account for the suppression of intermediate energy neutrinos. The
first one, neutrino oscillations in vacuum,
requires a large mixing angle and a seemingly unnatural
fine tuning of neutrino oscillation length with the Sun-Earth distance
 for intermediate energy neutrinos.
The second possibility, level-crossing effect
oscillations in presence of solar matter and/or
magnetic fields of regular and/or chaotic nature (MSW, RSFP), requires
 no fine tuning either for mixing parameter or neutrino mass difference
 to cause a selective large reduction of the neutrino flux. This mechanism
 explains naturally the suppression of intermediate energy neutrinos, leaving
 the low energy pp neutrino flux intact and high
energy $^8$B neutrinos only loosely suppressed.
Concrete range of parameters obtained including  the latest SK (Super-Kamiokande)  data will
be showed in the next section.

\subsection{The SK detector and Results.}

The  high precision and high statistics
 Super-Kamiokande (SK) experiment initiated operation in
 April 1996.
A few words about the detector itself.
SK is a 50-kiloton water Cerenkov detector
located near  the old Kamiokande detector under a
mean overburden of  2700 meter-water-equivalent.
The effective fiducial volume is  $22.5$ kt.
It is a well understood, well calibrated detector.
The accuracy of the absolute energy scale is estimated to be
$\pm 2.4\%$ based on several independent calibration sources:
cosmic ray through-going and stopping muons, muon decay
electrons, the invariant mass of $\pi^0$'s produced by neutrino interactions, radioactive source calibration,
and, as a novelty in neutrino experiments,
a 5-16 MeV electron LINAC.
In addition to the ability of recording higher
statistics in less time,
due to the much larger dimensions of the detector,
 SK can contain
multi-GeV muon events  making possible for the first
time a measurement of the spectrum of $\mu$-like events up to
$\sim 8-10 $ GeV.

The results from  SK, to be summarized below, combined with data from
earlier experiments provide important constraints on the MSW and
vacuum oscillation solutions for the solar neutrino problem (SNP), 
\cite{nu98xxx,smy99,SKnew}:

{\em Total rates.}
The most robust  results of the solar neutrino experiments so far are the total observed rates.
Preliminary results corresponding to the first 825 days of 
operation of SK (presented in spring'2000, \cite{SKnew}) with
a total number of events $N_{ev}= 11235\pm 180\pm 310$ in 
the energy range $E_{vis}=6.5-20$ MeV.
predict the following flux of solar ${}^8$B neutrinos:  
$$\phi_{{}^8 B}=(2.45\pm 0.04\pm 0.07)\times 10^6\ cm^{-2}\
 sec^{-1},$$
a flux which is clearly below the SSM expectations.
The most recent data on rates on all existing experiments are summarized in Table~(\ref{t1}).
Total rates alone indicate
 that the $\nu_e$ energy spectrum from the Sun is distorted.
The SSM flux predictions are inconsistent with the observed
rates in solar
neutrino experiments at approximately
the 20$\sigma$ level.
Furtherly, there is no linear combination of neutrino fluxes
that can fit the available data at the 3$\sigma$ level
[Fig.(\ref{SSM1}].
\TABLE[h]{
\centering
\small
\begin{tabular}{|c|c|c|c|}
 \hline\vspace{0.1cm}
Experiment      & Target & E. Th. (MeV)  & $S_{Data}/S_{SSM}\ (\pm 1\sigma)$
\\[0.1cm] \hline
SK-825d  & H$_2$O & $\sim$ 6.5-20  &$ 0.474\pm 0.020$\\
Homestake  & $^{37}$Cl & 0.8 &  $0.33\pm 0.03  $              \\[0.1cm]
Kamiokande & H$_2$O  & $\sim 7.5$ &   $0.54\pm 0.07$  \\[0.1cm]
SAGE       & $^{71}$Ga & 0.2 &                 $    0.52\pm 0.06 $    \\[0.1cm]
GALLEX     & $^{71}$Ga &0.2 &    $  0.60\pm 0.06 $  \\[0.1cm]
 \hline
\end{tabular}
\caption{Ratios of neutrino fluxes  by solar neutrino experiments to corresponding predictions from the  SSM
        (see Ref.\protect\cite{bah2}  and references therein, we take the INT normalization for the
   SSM data).}
\label{t1}
}

\begin{figure}[ht]
\centering
\psfig{file=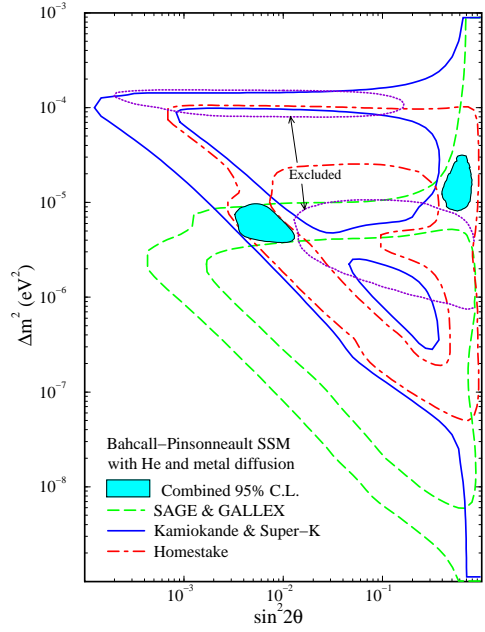,height=11cm} \vspace{-1cm} \caption{ The
result of the MSW parameter space (shaded regions) allowed by the
combined observations at 95\% C.L.\ aassuming the
Bahcall-Pinsonneault SSM with He diffusion.  The constraints from
Homestake, combined Kamiokande and Super-Kamiokande, and combined
SAGE and GALLEX are shown by the dot-dashed, solid, and dashed
lines, respectively.  Also shown are the regions excluded by the
Kamiokande spectrum and day-night data (dotted lines). [From Hata
and Langacker, Ref.(\protect\cite{hata} and references therein.]}
\label{SOLAR1}
\end{figure}

{\em Zenith angle: day-night effect.}
If MSW oscillations are effective, for a certain range of  neutrino
parameters the observed event rate will depend upon the
zenith angle of the Sun (through a Earth matter regeneration effect).
Win present statistics, the most
 robust  estimator of zenith angle dependence  is the day-night
 (or up-down) asymmetry, A. The experimental estimation is \cite{SKnew}:
\begin{eqnarray}
A\equiv\frac{N-D}{N+D}=0.032\pm0.015\pm 0.006, \quad (E_{recoil}>6.5\ {\rm MeV}).
\end{eqnarray}
The difference is small and not statistically significant
but it is in the direction that would be expected from
regeneration at Earth (the Sun is apparently neutrino brighter at night).
Taken alone the small value observed for A excludes a large part of
the  parameter region  that is allowed if only the
total rates would be considered [see Fig.(\ref{SOLAR1})].

\begin{figure}[ht]
\centering
\psfig{file=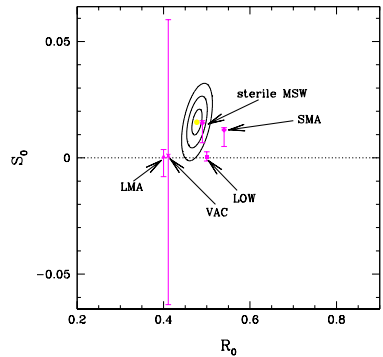,height=6cm} \vspace{-0.5cm}
\caption[]{Deviation from an undistorted energy spectrum. The
$1,2,3\sigma$ allowed regions are shown  in the figure. The ratio
of the observed counting rate as a function of electron recoil
energy~\protect\cite{nu98xxx} to the expected undistorted energy
spectrum was fit to a linear function of energy, with intercept
$R_0$ and slope $S_0$. The five oscillation solutions SMA active
and sterile, LMA, LOW, and vacuum oscillations,
 all provide
acceptable fits to the data, although the fits are not
particularly good. (From Bahcall and Krastev,
Ref.\protect\cite{bah2}).} \label{SOLAR2}
\end{figure}

{\em Spectrum Shape.}
The shape of the
 neutrino spectrum determines the shape of the recoil electron
 energy spectrum produced by neutrino-electron scattering 
in the detector and is independent of the astrophysical source.
All the
neutrino oscillation solutions (SMA,LMA,LOW and Vacuum) provide
acceptable, although not excellent fits to the recoil
energy spectrum. The simplest test is to investigate whether the
ratio, R, of the observed to the standard energy spectrum is a
 constant with increasing energy. 
The   null flatness hypothesis  is accepted at the 
90\% CL ($\chi^2\sim 1.5$, \cite{SKnew}). However, alternative
 fits of the ratio $R$ to a linear function of energy  yields
 slope values does not discard  the presence of distortion 
at higher energies [see Figs.(\ref{SOLAR1}-\ref{SOLAR2}) and 
next paragraph].

\begin{figure}[ht]
\centering
\psfig{file=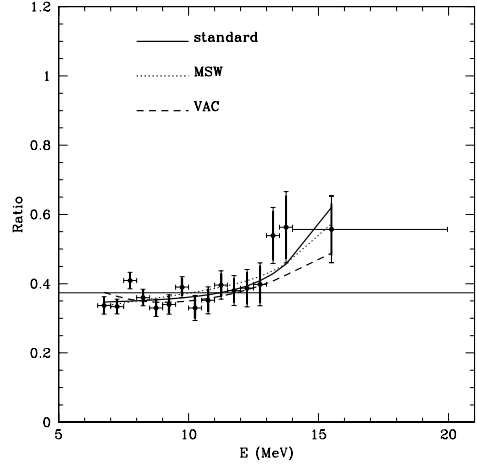,height=7cm} \vspace{0cm}
\caption{ Nuclear physics calculations of the rate of the hep
reaction are uncertain. The  figure  show the results for the
predicted energy spectrum that is measured by SK
(\protect\cite{nu98xxx}). The total flux of hep neutrinos was varied
to obtain the best-fit for each scenario. The calculated curves
are global fits to all of the data, the chlorine, GALLEX, SAGE,
and SK total event rates, the SK energy spectrum and Day-Night
asymmetry. (Figure reproduced from Ref.\protect\cite{bah5}).}
\label{SOLAR6}
\end{figure}

{\em Spectrum shape: the hep neutrino problem.}
A small but significant discrepancy appears
when comparing  the predictions from the
global best fits for the  energy spectrum
at high energies
($E_{\nu}\gsim 13 $ MeV)
with the SK results.
From this discrepancy it has been speculated that
uncertainties on the $hep$ neutrino fluxes  
may affect the higher energy solar neutrino energy spectrum.
Presently
low energy nuclear physics calculations of the rate of 
the hep reaction are  uncertain by a factor of, at least, six.
Coincidence between expected and measured ratios is improved
when the hep flux is  allowed to vary
as a free parameter [see Fig.(\ref{SOLAR6}) and Ref.\cite{SKnew}]. The best fit is obtained by a combination $\phi\sim 0.45 {}^8\mbox{ B}+16 \mbox{hep}$ ($\chi^2\sim 1.2$. An upper limit on the  ratio of experimental to SSM $hep$ flux is obtained:
$$\phi_{hep}^{exp}/{ \phi_{hep}^{BP98}}< 15, \ (90\% CL).$$

{\em Seasonal Variation.}
No evidence for a anomalous seasonal variation of the neutrino flux has 
been found. The  results (SK 825d, $E_{vis}=10-20$ MeV)
 are consistent with what is expected 
from a geometrical variation due to the Earth orbital eccentricity ($\chi^2\sim 0.5$ for the null hypothesis, Ref.\cite{SKnew}).

{\em Analysis of data.}
{}From a two-flavor  analysis (Ref.\cite{SKnew}, see also Ref.\cite{bah2,bahcall99}) of the total event rates in the
ClAr, SAGE,GALLEX and SK experiments the best $\chi^2$ fit considering
 active neutrino oscillations is obtained for
$\Delta m^2=5.4\times 10^{-6}$ eV$^2, \sin^2 2\theta=5.0\times 10^{-3}$
(the so called small mixing angle solution, SMA).
Other local $\chi^2$ minima exist. 
The large mixing angle solution
(LMA) occurs at
$\Delta m^2=3.2\times 10^{-5}$ eV$^2, \sin^2 2\theta=0.76$,
the LOW solution (lower probability, low mass), at
$\Delta m^2=7.9\times 10^{-8}$ eV$^2, \sin^2 2\theta=0.96$.
The vacuum oscillation solution occurs at
$\Delta m^2=4.3\times 10^{-10}$ eV$^2, \sin^2 2\theta=0.79$.
At this extremely low value for the  mass difference the MSW effect is 
inopperant.

For oscillations involving sterile neutrinos
(the  matter effective potential is modified in this case)
the LMA and LOW solutions are
not allowed and only the  (only slightly modified)
SMA solution together with the vacuum solution  are still possible.

In the case where all data, the total rates, the zenith-angle
dependence and the recoil energy spectrum,
is combined the best-fit solution is almost identical
to what is obtained for the rates-only case.
For other solutions, only the SMA and  vacuum solution 
survives (at the 99\% CL).
The LMA and the LOW solutions are, 
albeit marginally, ruled out \cite{bah2}.

{\em Solar magnetic Fields and antineutrino flux bounds.}
Analysis which  consider  neutrino propagation in
presence of solar magnetic fields
have also been presented. In this case a variant, more complicated, version of the MSW effect, the so called RSFP effect could manifest itself. Typically, these analysis
 yield   solutions with
$\dms\sim 10^{-7}-10^{-8}$ eV$^2$ for both small and 
large mixing angles.
Spin flavor or resonant spin flavor (RSFP)  solutions are much more ambiguous than pure MSW solutions because of
necessity of  introducing additional  
free parameters in order to model
 the largely unknown intensity and profile of 
solar magnetic field.
The recognition of the random nature of solar 
convictive  fields
and recent theoretical developments in the treatment of Schroedinger
random equations have
partially improved this situation, allowing the obtaining of
SNP solutions without the necessity of a 
detailed model description
 (see recent analysis in \cite{tor2a,tor2b,tor2c,bykov,tor5}).

In addition, random RSFP models predict the production of
a sizeable quantity of electron antineutrinos in case the neutrino is a Majorana particle.

Presently, antineutrino searches \cite{tor6} with negative
results in Kamiokande and SK   are welcome because 
restrict significantly
the, uncomfortably large, parameter space of RSFP models.

A search \cite{tor6} for inverse beta decay electron antineutrinos has set
limits on the absolute flux of solar antineutrinos originated 
from the solar ${}^8$B neutrino component:
$$\Phi_{\overline{\nu}}({}^8 B)< 1.8\times 10^5 \mbox{cm}^{-2} \mbox{s}^{-1},\ (95\% \ \mbox{CL}), $$
a number which is equivalent to an averaged conversion probability bound of (with respect the SSM-BP98 model)
$$P<3.5\%\ ( 95\% \ \mbox{CL}).$$
 
\begin{figure}[ht]
\centering \epsfig{file=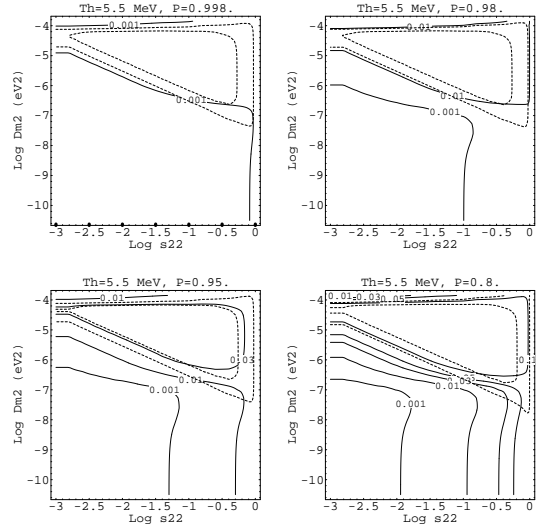,height=8cm}
\vspace{0cm} \caption{ Solid lines: antineutrino production
probability as a function of the neutrino mixing angle $\log\
\sin^2 2\theta$ and $\Delta m^2$. Dashed thick lines: $S/S_{SSM}$
signal rates at SK (contours $=0.3,0.5$). The threshold neutrino
energy is in this case: $E_{th}=5.5$ MeV. From left to right and
from top to bottom: $P=0.998,0.98,0.95,0.8,0.7,0.5$. The
corresponding  r.m.s fields  are
 $\surd{\langle \tilde{B}^2\rangle}=15,45,65,150,220$ and 600 kG
respectively  (supposing the scale $L_0=1000$ Km and
$\mu=10^{-11}\mu_B$). (From  Ref.\protect\cite{tor2c})}
\label{efig}
\end{figure}

In the future such antineutrinos
 could be identified both in SK or in SNO experiments
 setting the Majorana nature of the neutrino. 
In Ref.\cite{tor2c} [see Fig.(\ref{efig}) for illustration]
 it has been shown that, even for moderate levels of
noise, it is possible to obtain a
probability for    $\nu_e\to \overline{\nu}_e$
 conversions about $\sim 1-3\%$
in the energy range 2-10 MeV  for large regions of
the mixing parameter space while still satisfying present
SK antineutrino
bounds and observed total rates.
In the other hand it would be  possible
to obtain information about the
solar magnetic internal field if
antineutrino bounds reach the $1\%$ level and
a particle physics
solution to the SNP is assumed.


\subsection{Atmospheric neutrinos}

Atmospheric neutrinos are the decay products of hadronic
showers produced
 by cosmic ray interactions in the atmosphere.
The composed  ratio R
$$R\equiv \left ( \mu/e\right )_{DATA}/\left ( \mu/e\right )_{MC} $$
where $\mu/e$ denotes the ratio of the number of $\mu$-like to
$e$-like neutrino interactions observed in the experiment or predicted by the simulation
is  considered as an 
 estimator of the atmospheric neutrino
flavor ratio
$(\nu_\mu+\overline{\nu}_\mu)/(\nu_e+\overline{\nu}_e).$
The calculations of individual  absolute neutrino fluxes
  have large uncertainties  at the  $\sim 20\%$ level \cite{atmfluxes}.
However, the flavor flux ratio  is known
 to an accuracy of better than
$5\%$ in the energy range \mbox{$0.1-10$} GeV.
The calculated flux ratio has a value
of about 2 for energies $<$ 1 GeV and increases with increasing neutrino energy reaching a value $\sim 10$ at $100$ GeV.
The angle distribution of the different fluxes is also an 
important ingredient in the existing evidence for atmospheric 
neutrino oscillations. Calculations show that 
for neutrino energies
higher than a few GeV, the fluxes of upward and
downward going neutrinos are expected to be nearly equal;
geomagnetic field
 effects at these energies are expected to be small because of the relative
large geomagnetic
rigidity of the primary cosmic rays that produce these
neutrinos \cite{atmfluxes}.

Prior to the present era dominated by Super-Kamiokande results,
anomalous, statistically significant,
low values of the ratio $R$  have 
 been repeatedly obtained previously
 \cite{experiments,experiments2}
 in the water Cerenkov detectors Kamiokande and IMB-3
and in the calorimeter-based   Soudan-2  experiment 
for ``sub-GeV'' events (E$_{vis}< 1$ GeV).
The NUSEX and Frejus experiments  
reported however results
consistent with no deviation from unity 
with  smaller data samples.
Kamiokande  experiment 
observed a value of $R$ smaller than unity
in the  multi-GeV (E$_{vis}>$1 GeV) energy region  as well as a
dependence of this ratio on the zenith angle.
IMB-3, with a smaller data sample,
 reported inconclusive results in a similar energy
 range, not in contradiction with Kamiokande observations \cite{experiments,experiments2}.

Super-Kamiokande (SK)  results are completely consistent
with previous results at a much higher accuracy level.
Specially significant improvements in accuracy have been
 obtained in measuring the zenith angular dependence of the 
neutrino events: in summary, the single most significant 
 result obtained by SK
is that the flux of muon neutrinos going up 
is smaller than
that of down-going neutrinos.

As we commented before,
in addition to the ability of recording higher
statistics in less time,
due to the much larger dimensions of the detector, the SK  
detector can contain
multi-GeV muon events  making possible for the first
time a measurement of the spectrum of $\mu$-like events up to
$\sim 8-10 $ GeV.
{}From experimental and phenomenological reasons,
the SK experiment uses the following 
event classification nomenclature.
 According to their origin, 
events can be classified as  
{\em e-like} (showering, $\nu_e$ or $\overline{\nu}_e$ events) or
{\em $\mu$-like} (non-showering, 
$\nu_\mu$ or $\overline{\nu}_\mu$ events).
According to the position of the neutrino interaction, they 
distinguish 
{\em contained events}
 (vertex in fiducial volume, $98\%$ muon induced), 
which, depending on their energy, are typed as 
{\em sub-GeV} ($E<\sim 1$ GeV) or 
{\rm multi-GeV samples} ($E<\sim 10$ GeV).
Non-contained events can be:
{\em Upward through-going muons} 
(vertex outside the detector, muon induced, $E_\nu\sim 500$ GeV)
or
{\em Upward stopping muons} (typically $E_\nu<\sim 50 $ GeV). 
In all cases, 
the neutrino path-length covers the full range, from 
 $\sim 10^1$ km for {\em down} events
 to $10^4$ km for {\em up} events.
In what follows we summarize the present results about 
 total and zenith-angle dependent rates.

\begin{figure}[ht]
\centerline{\protect\hbox{\psfig{file=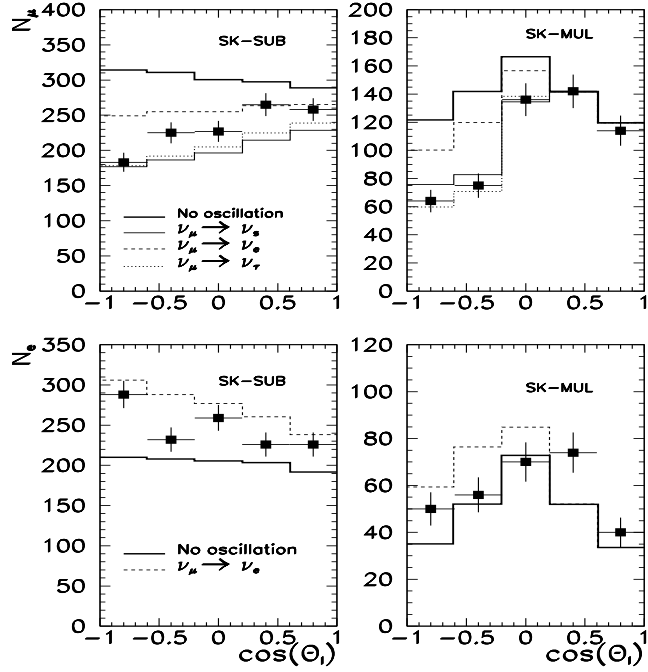,width=0.6\textwidth,height=0.4\textheight}}}
\caption{Angular distribution for Super-Kamiokande electron-like
and muon- like sub-GeV and multi-GeV events. Predictions in the
absence of oscillation (thick solid line), $\nu_\mu \to \nu_s$
(thin solid line), $\nu_\mu \to \nu_e$ (dashed line) and $\nu_\mu
\to \nu_\tau$ (dotted line).  The errors displayed in the
experimental points is only statistical. (From
Ref.\protect\cite{atm1})} \label{ATM8}
\end{figure}

{\em Total rates.}
In the sub-GeV range ($E_{vis}< 1.33$ GeV), 
{}From an exposure of  61 kiloton-years (kty) (990 days of operation) 
of the SK detector the measured ratio $R$ is:
$$R_{sub gev}=0.66\pm0.02\pm 0.05. $$
It is not possible
 to determine from data, whether the observed deviation of $R$ is due to an electron excess of a muon deficit.
The distribution of $R$ with momentum in the
sub-GeV range is consistent with a flat distribution
within the statistical error as  happens
with zenith angle distributions [see right plots in Fig.(\ref{ATM8})].

In the multi-GeV range, it has been obtained (for a similar
exposure) a ratio $R$ which is
slightly higher than at lower energies 
$$R_{multi gev}=0.66\pm0.04\pm 0.08. $$
For e-like events, the data is apparently consistent with
MC. For $\mu$-like events there is a clear discrepancy between
 measurement and simulation.

\begin{figure}[ht]
\centering
\epsfig{file=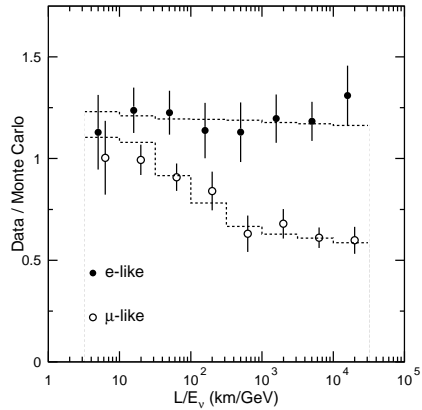,height=6cm} \vspace{0cm} \caption{The SK
multi-GeV data sample. The ratio of the number of FC (fully
contained) data events to FC Monte Carlo events versus
reconstructed $L/E_\nu$. Points: absence of oscillations. Dashed
lines: expected shape for $\numu \leftrightarrow \nutau$ at
$\dms=2.2\times10^{-3} $eV$^2$ and $\sstt=1$. The slight $L/E_\nu$
dependence for $e$-like events is due to contamination (2-7\%) of
$\nu_\mu$ CC interactions. (From Ref.\protect\cite{sk2}).}
\label{ATM4}
\end{figure}

{\em Zenith Angle.}
A strong distortion in the shape of the $\mu$-like event
 zenith angle distribution was observed
[Plots (\ref{ATM4}-\ref{ATM8})].
The angular correlation between the neutrino direction
and the produced
charged lepton direction is much better at higher energies ( $\sim 15^0-20^0$):
the zenith angle distribution of leptons reflects 
rather accurately that of the neutrinos in this case.

At lower energies, the ratio of the number of
upward to downward $\mu$-like events was found to be
$$(N_{up}/N_{down})^\mu_{Data}=0.52\pm 0.07$$
while the expected value is practically one:
$$(N_{up}/N_{down})^\mu_{MC}=0.98\pm 0.03.$$
The validity of the results  has been  tested by
measuring the azimuth angle distribution of the
incoming neutrinos, which is insensitive to
a possible influence from neutrino oscillations.
This shape agreed with MC predictions which were nearly flat.

Another signal for the presence of neutrino oscillations could 
be present in the ratio of neutrino events for two well separated 
energy ranges. This is the case for the ratio between 
 upward through going to upward stopping muon events, both classes 
 correspond to very high energy events. 
The results and expected values are the following 
(\cite{SKmoriond2000,atmfluxes})
\begin{eqnarray}
\left (N_{stop}/N_{throug}\right )_{Data}^\mu& =&
0.23\pm0.02 \\
\left (N_{stop}/N_{throug}\right )_{MC}^\mu &=& 0.37\pm 0.05.
\end{eqnarray}
The ratio of data to MC is $\sim 0.6$. With these results,  
the probability that they do correspond to  no-oscillation
 scenario is rather low, $P\sim 10^{-4}-10^{-3}$ 
\cite{SKmoriond2000}.

{\em Analysis.} Oscillation parameters are measured by 
several samples (FC, PC, up-stop, up-through). The result is 
that all samples are overall consistent with each other.
This hypothesis fits well to the  angular distribution,
since  there is a large
difference in the neutrino path-length between upward-going
($\sim 10^{4}$ Km) and downward-going ($\sim 20$ Km): a zenith
angle dependence of $R$ can be interpreted as  a clear-cut 
evidence  for
neutrino oscillations.

\begin{figure}[ht]
\centering
 \epsfig{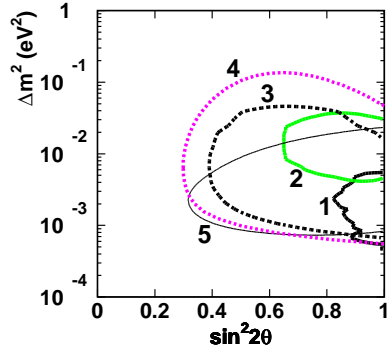}
\caption{ The allowed neutrino oscillation parameter regions
obtained by Kamiokande and SK ( 90\% C.L..).
 (1) and (2):the regions obtained by contained event analyses from
Super-Kamiokande and Kamiokande, respectively. (3) and (4):
upward through-going muons from SK and Kamiokande, respectively.
(5) stopping/trough-going ratio analysis of upward going muons
from SK. (From  Ref.\protect\cite{atm2})} 
\label{ATM6}
\end{figure}

Among the different possibilities, the 
most obvious solution to the observed discrepancy is
 $\nu_\mu\to\nu_\tau$ flavor neutrino oscillations.
$\nu_\mu-\nu_e$ oscillations does not fit however so well, they would also
conflict laboratory measurements [CHOOZ, see figs.(\ref{fCHOOZ}-\ref{ATM6})].

Oscillation into sterile neutrinos, $\nu_\mu\to \nu_s$,
could also be in principle  a good explanation consistent with data.
Different tests has been performed for  distinguishing 
$\nu_\mu\to\nu_\tau$  from 
$\nu_\mu\to\nu_s$  oscillations:
A possible test of $\nu_\mu\to \nu_s$ vs $\nu_\mu\to \nu_\tau$ oscillations is provided by the study of the $\pi^0/e$ ratio \cite{nakahata}. 
In the $\mu-\tau$ case, the $\pi^0$ production  due to neutral current
interactions do not change, causing  the  $\pi^0/e$ ratio to be the 
same as the expectation without neutrino oscillations. In the sterile case
 such a ratio should be smaller ($\sim$ 83\%) 
than expected because the absence of 
$\nu_s$ neutral current interactions. 
 $\pi^0$ events 
experimental identification can be performed by study of 
 their invariant mass distributions  compared with 
 Monte Carlo simulations. 
Present results conclude that 
the $\nu_\mu\to\nu_s$ oscillation hypothesis is 
disfavored  at the  $99\% CL$.

Evidence for oscillations equals
evidence for   non-zero neutrino mass within the standard neutrino
theory.
The allowed neutrino oscillation parameter regions
obtained by Kamiokande and SK  from different analysis are
shown in Fig.(\ref{ATM6}).
Under the interpretation 
as  $\nu_\mu\to\nu_\tau$ oscillations, the best fit provide
    $\Delta m^2=\sim 2-5 \times 10^{-3}$ and a very 
large mixing angle $\sin^2 2\theta> 0.88$.
Unless
there is no fine tuning, this suggests  a
neutrino mass of the order of 0.1 eV.
Such a mass implies the neutrino energy density in
the universe to be 0.001 of the critical density
which is too small to have cosmological consequences.
This is of course a very rough argument:
specific models, however, may allow larger neutrino masses
quite naturally.


\subsection{Global multi-fold analysis and the necessity for
sterile neutrinos.}

{}From the individual  analysis  of the data available from neutrino
experiments, it follows that there exist three different
scales of neutrino mass squared differences and two different
ranges of small and maximal mixing angles, namely:
\begin{eqnarray}
\Delta m_{sun}^2&\sim  10^{-5}-10^{-8}
\ eV^2\ ,& \sin^2 2\theta\sim 7\times 10^{-3}
(MSW,RSFP), \\
&\sim 10^{-10} \ eV^2,& \sin^2 2\theta\sim 0.8-0.9\ (Vac.); \\
\Delta m_{Atm}^2&\sim  5\times 10^{-3} \ eV^2,&\  \sin^2 2\theta\sim 1\\
\Delta m_{LSND}^2&\sim  3\times 10^{-1}-2\ eV^2&\  \sin^2 2\theta\sim 10^{-3}-10^{-2}.
\label{e1001}
\end{eqnarray}
Fortunely for the sake of simplicity the neutrino mass scale relevant
 for HDM is roughly similar to the LSND one. The  introduction of the
former  would not change any further conclusion.
But for the same reason, the definitive refutation of LSND results by
KARMEN or future experiments does not help completely in simplifying the
task  of finding a consistent framework for all the
neutrino phenomenology.

Any combination of experimental data which involves only of the
two mass scales can be fitted within a three family scenario, but
solving simultaneously the solar and atmospheric problems requires generally
some unwelcome fine tuning of parameters at the $10^{-2}$ level.
The detailed analysis of Ref.\cite{BILENKY}
 obtains for example that
 solutions with 3 neutrino families which are compatible with the
results from SBL inclusive experiments, LSND and solar neutrino experiments are possible.
Moreover it has been shown that it is possible to obtain, under
simple assumptions but without a detailed fit of all possible parameters,
very concrete expressions for
the $3\times 3$ mixing matrix, see for example the early
 Ref.\cite{tor-quasi}, of which the called bi-maximal model
of Ref.\cite{bimaximal}  is a particular case.
The real problem arises when
 one add the results from CHOOZ, which rule out large
atmospheric $\nu_\mu \nu_e$ transitions and zenith dependence from
SK atmospheric data one comes to the necessity of
consideration of schemes with four massive neutrinos
  including  a light sterile neutrino.
Among the numerous possibilities,
 complete mass hierarchy of four neutrinos is not favored by existing
data \cite{BILENKY} nor four-neutrino mass spectra with one
neutrino mass separated from the group of the three close masses by the
''LSND gap'' ($\sim$ 1  eV). One is left with two possible
 options where  two double-folded groups of close masses
are separated by a $\sim 1 $ eV gap:
\begin{eqnarray}
&(A)&\ \underbrace{\overbrace{\nu_e\to\nu_s: \ m_1< m_2}^{sun}<< \overbrace{\nu_\mu\to \nu_\tau:\ m_3< m_4}^{atm}}_{LSND\sim 1 eV} \\[0.1cm]
&(B)&\ \underbrace{\overbrace{\nu_e\to \nu_\tau: \ m_1< m_2}^{sun}<< \overbrace{\nu_\mu\to\nu_s:\ m_3< m_4}^{atm}}_{LSND\sim 1 eV}.
\end{eqnarray}

The two models would be distinguishable from the detailed
analysis of future
 solar and atmospheric experiments.
For example they may be tested
combining future precise  recoil electron spectrum in $\nu e\to \nu e$
measured in SK and SNO ( see Ref.\cite{SNO}  for experiment details and
Refs.\cite{bah10} for performing expectations) with
the  SNO spectrum measured in CC absorption.
The SNO experiment
(a 1000 t heavy water under-mine  detector)
will measure the rates of the charged (CC) and neutral (NC) current reactions
induced by solar neutrinos in deuterium:
\begin{eqnarray}
&& \nu_e + d \rightarrow p+p+e^-\quad({\rm CC\ absorption})\nonumber \\
&& \nu_x + d \rightarrow p+n+\nu_x\quad({\rm NC\ dissociation}).
\label{reactionNC}
\end{eqnarray}
including the determination of the electron recoil energy in the CC
reaction. Only the more energetic $^8$B solar neutrinos
are expected to be detected since the expected
SNO threshold  for CC events is an electron kinetic energy
of about 5 MeV and the physical threshold for NC dissociation is the
binding energy of the deuteron, $E_b= 2.225$ MeV.
If the (B) model it is true one
expects $\phi^{CC}/\phi^{NC}\sim 0.5$ while in the
(A) model the ratio would be $\sim 1$.
The schemes (A) and (B) give different predictions for the neutrino mass
 measured in tritium $\beta$-decay and for the effective Majorana mass
 observed in neutrinoless double $\beta$ decay. Respectively we have
$\mid \langle m\rangle\mid < m_4$ (A) or $<< m_4$ (B).
Thus, if scheme (A) is realized in nature this kind of experiments can see the
effect of the LSND neutrino mass.

{}From the classical LEP requirement $N_\nu^{act}=2.994\pm 0.012$ \cite{PDG98}, it is clear that
the fourth neutrino  should be  a $SU(2)\otimes U(1)$ singlet in order
to ensure that does not affect the invisible Z decay width.
The presence of additional weakly interacting light particles, such as a
light sterile $\nu_s$, is constrained by BBN since it would enter into
equilibrium with the active neutrinos via neutrino oscillations (see Section~\ref{sectioncosmos}).
The limit
$\Delta m^2 \sin^2 2\theta< 3\times 10^{-6}$ eV$^2$ should be
fulfilled in principle.
However systematical uncertainties in the
derivation of the BBN bound make any bound too unreliable to be
taken at face value and can eventually be avoided \cite{foot}.
Taking the most restrictive options
(giving $N_\nu^{eff}< 3.5$) only the  (A) scheme is allowed, one where the
sterile neutrino is mainly mixed with the electron neutrino.
In the lest restrictive case ($N_\nu^{eff}< 4.5$)
both type of models would be allowed.

\section{Conclusions and future perspectives.}

The theoretical challenges that  the present phenomenological situation
 offers are two at least: to understand origin and,
very particularly, the lightness of the sterile
neutrino (apparently requiring a radiatively generated mass) and
to account for the maximal neutrino mixing indicated by the
atmospheric data which is at odd from which one could expect from
considerations of the mixing in the quark sector.
Actually, the existence of light sterile neutrinos could even be
beneficial in diverse astrophysical and cosmological scenarios (supernova
nucleosynthesis, hot dark matter, lepton and baryon asymmetries for example).

In the last years different indications in favor of nonzero neutrino masses
and mixing angles have been found.
These evidences include
 four solar experiments clearly
demonstrating an anomaly compared  to the predictions of the
Standard Solar Model (SSM) and a number of other atmospheric
experiments, including a high statistics, well calibrated one, demonstrating
 a quite different anomaly at the Earth scale.

One could argue that if we are already  beyond the stage of having only
''circumstantial evidence for new physics'', we are still however a long
way from having  ''conclusive proof of new physics''. Evidence for new
physics does not mean the same as evidence for neutrino oscillations
but there exists  a significant  case for neutrino oscillations and hence
neutrino masses and mixing  as ''one'', indeed the most serious candidate,
explanation of the data.

Non-oscillatory alternative explanations of the neutrino anomalies are also
possible but any of them will not be specially elegant or economical
(see Ref.\cite{pakvasa} for a recent summary and references therein):
they will involve anyway non-zero neutrino masses and mixing.
As a result even if neutrinos have masses and do mix, the observed neutrino
anomalies,  may  be a manifestation of a complicated mixture of effects due to
oscillations and effects due to  other exotic new physics. The dominant effect
 would not necessarily be the same in each energy or experimental domain.
The list of  effects due to exotic physics which have
been investigated in some degree in the literature, would include
\cite{pakvasa}:
Oscillation of massless neutrinos via FCNC and Non-Universal neutral
currents (NUNC) which has been considered as feasible
explanation for the solar neutrino
observations \cite{exotic1} and atmospheric neutrinos \cite{exotic3}.
It has also been studied the possibility of decaying neutrinos as possible
solutions in the solar case \cite{exotic2} and in the atmospheric case
\cite{exotic4}.
LSND results could be accounted for without oscillations provided
that muon conventional decay modes are accompanied by rare modes  including
standard and/or sterile neutrinos. In this case the energy or distance
dependence, typical of the oscillation explanation, would be absent
\cite{exotic0}.

Finally, explanations of the neutrino experimental data which
involve alterations of the basic framework of known physics (quantum and
relativity theory) have been proposed:
 A similar signature to neutrino decay would be produced by a
huge, non-standard, quantum decoherence rate along the neutrino
propagation \cite{exotic5}.
Proposed explanations involving  relativity effects include gravitationally
induced oscillations (see, for example, Ref.\cite{fogligrav}) or violation of
Lorenz invariance \cite{exotic6}.
In the first case it has been suggested that different flavors have different coupling to the gravitational potential.
In the second case it is claimed the  existence of different  maximum speeds
for each neutrino specie. In both cases, rather than the usual dependence
on $L/E$, one finds a $L\times E$ as a characteristical signature.

Of course, one possible alternative is that one or more of the
experiments will turn out to be wrong. This is possible, probable and
 even desirable from a phenomenologist point of view because
his/her task would be considerably simplified as we have seen above.
What  it is little probable, with all the evidence accumulated by now, is that
 all the experiments turn out to be simultaneously wrong.

\begin{figure}[ht]
\centering
\psfig{file=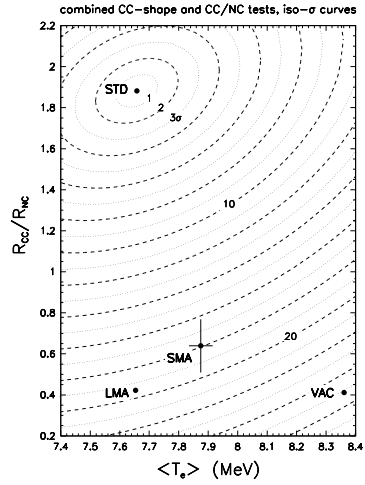,height=8cm} \vspace{0cm}
\caption{Conclusive probes of lepton number violation in solar
neutrino experiments. Iso-sigma contours for the SNO for the
combined CC-shape and CC/NC test, for the representative
oscillation cases. Iso-sigma contours
 for the combined CC-shape and CC/NC test, for
representative oscillation cases. STD = standard (no oscillation);
SMA = small mixing angle (MSW); LMA = large mixing angle (MSW);
VAC = vacuum oscillation. (From  Ref.\protect\cite{bahcall2})}
\label{FUT3}
\end{figure}

Many neutrino experiments are taking data, are going to start or are
under preparation: solar neutrino experiments (SNO and Borexino are of
major interest, also HERON, HELLAZ, ICARUS, GNO and  others);
LBL reactor (CHOOZ, Palo Verde, KamLand) and accelerator experiments
(K2K, MINOS, ICARUS and others);
SBL experiments  (LSND, KARMEN, BooNe and many others).
The important problem for any  next generation experiment is to find
specific and unambiguous
experimental probe that the ''anomalies'' which has been found
 are indeed signals of neutrino oscillations and to distinguish among the
different neutrino oscillation possibilities (this is specially important
in the Solar case).
Among these probes, we could include:
\begin{itemize}
\item Perhaps the  most direct test of
SM deviation:   to measure the
ratio of the flux of $\nu_e$'s (via a  CC interaction) to the
flux of neutrinos of all types ($\nu_e + \nu_\mu + \nu_\tau$,
determined
by  NC interactions). This measurement will be
done hopefully by the SNO experiment in the near future [see Fig.(\ref{FUT3})].

\item Statistically significant demonstration of an
 energy-dependent modification of the shape of the electron neutrino
spectrum arriving at Earth. Besides observing distortion in the shape of
$^8$B neutrinos, it will be very important to make direct
measurements of the $^7$Be (Borexino experiment)
and pp (HERON,HELLAZ) neutrinos.
\item Improved observation of a zenith angle effect
in atmospheric experiments or their equivalent,  a
 day-night effect  in solar experiments.
\item And least, but by no means the least, independent confirmation by
one or more accelerator experiments.
\end{itemize}
There is a high probability that in the near future  we should know much more
than now about the fundamental properties of neutrinos and their
 masses, mixing and their own nature whether Dirac or Majorana.

\vspace{1cm}
{\bf Acknowledgments}

The authors are supported by research grants from the 
Spanish Ministerio de Educacion y
Cultura.

\listoftables

\listoffigures

\newpage
\section*{Figures}

\end{document}